\documentclass[twocolumn,journal, bookmarks=false]{IEEEtran}
\IEEEoverridecommandlockouts
\usepackage{color}
\usepackage{graphicx}
\usepackage{amsmath}
\usepackage{amssymb}
\usepackage{algorithm}
\usepackage{algorithmic}
\usepackage{amsmath}
\usepackage{multirow}
\usepackage{booktabs}
\usepackage{array}
\usepackage{amsthm}
\usepackage{stfloats}
\usepackage{caption}
\usepackage{subfigure}
\usepackage{epstopdf}
\usepackage{bm}
\usepackage{setspace}
\usepackage{url}

\newcommand{\be}{\begin{equation}}
\newcommand{\ee}{\end{equation}}
\newcommand{\bea}{\begin{eqnarray}}
\newcommand{\eea}{\end{eqnarray}}
\newcommand{\ba}{\begin{array}}
\newcommand{\ea}{\end{array}}
\newcommand{\nid}{\noindent}
\newcommand{\non}{\nonumber}

\usepackage{flushend}

\allowdisplaybreaks[4]

\title{Target Detection in ISAC Systems with Active RISs: A Multi-Perspective Observation Approach
\thanks{S. Zhang, M. Li, and W. Wang are with the School of Information and Communication Engineering, Dalian University of Technology, Dalian 116024, China (e-mail: zhangshoushuo@mail.dlut.edu.cn; mli@dlut.edu.cn; wangwei2023@dlut.edu.cn).}
\thanks{R. Liu is with the Center for Pervasive Communications and Computing, University of California, Irvine, CA 92697, USA (e-mail: rangl2@uci.edu).}
\thanks{Q. Liu is with the School of Computer Science and Technology, Dalian University of Technology, Dalian 116024, China (e-mail: qianliu@dlut.edu.cn).}
}

\author{Shoushuo Zhang,
        Rang Liu,~\IEEEmembership{Member,~IEEE,}
        Ming Li,~\IEEEmembership{Senior Member,~IEEE,}
        Wei Wang,~\IEEEmembership{Senior Member,~IEEE,}
        and Qian Liu,~\IEEEmembership{Member,~IEEE}
        }

\begin{document}

\maketitle

\pagestyle{empty}
\thispagestyle{empty}

\begin{abstract}
Integrated sensing and communication (ISAC) has emerged as a transformative technology for 6G networks, enabling the seamless integration of communication and sensing functionalities. Reconfigurable intelligent surfaces (RIS), with their capability to adaptively reconfigure the radio environment, have shown significant potential in enhancing communication quality and enabling advanced cooperative sensing. This paper investigates a multi-RIS-assisted ISAC system and introduces a novel multi-perspective observation framework that leverages the diversity of multiple observation paths, each exhibiting distinct spatial, delay, and Doppler characteristics for both target and clutter.
The proposed framework integrates symbol-level precoding (SLP) and space-time adaptive processing (STAP) to fully exploit the benefits of multi-perspective observations, enabling superior target-clutter separation and significantly improving detection accuracy. The objective is to jointly design the transmit waveform, reflection coefficients of multiple active RISs, and spatial-temporal receive filters to maximize the radar output signal-to-clutter-plus-noise ratio (SCNR) for target detection, while ensuring the quality-of-service (QoS) requirements of communication users. To address the resulting non-convex optimization problem, an effective iterative algorithm is developed, combining fractional programming (FP), majorization-minimization (MM), and the alternating direction method of multipliers (ADMM). Extensive simulation results validate the effectiveness of the proposed multi-perspective observation strategy, demonstrating its advantages in improving target detection performance in challenging environments.
\end{abstract}

\section{Introduction}\label{sec:introduction}

Integrated sensing and communication (ISAC) is emerging as the cornerstone technology for next-generation mobile networks. By enabling the shared use of spectrum and hardware resources, ISAC facilitates the simultaneous realization of reliable communication and precise sensing tasks. This dual functionality enhances spectrum efficiency while substantially reducing system costs and power consumption, representing a critical step toward more sustainable and cost-effective communication systems \cite{LiuFan JSEC 2022}-\cite{Liu Fan TCOM 2020}. Extensive research on ISAC systems has addressed various aspects, including beamforming design, waveform optimization, resource allocation, interference management, and advanced signal processing techniques \cite{Liu Fan TSP 2022}-\cite{Liu Rang JSEC 2022}. These studies highlight that efficient system design and signal processing strategies can effectively mitigate interference between communication and radar functionalities, enabling mutual benefits and seamless integration.

Existing studies on ISAC systems that rely solely on direct-path sensing face inherent limitations \cite{Liu Guangyi JSEC 2024}-\cite{Wei Zhiqing 2024}. The direct-path approach is highly vulnerable to interference from environmental scatterers and obstructions, such as buildings, making it unreliable for robust sensing applications \cite{Liu Guangyi JSEC 2024}. Additionally, the single-perspective observation inherent in direct-path sensing often results in limited target information, leading to positioning inaccuracies and constraints in velocity estimation \cite{Wei Zhiqing 2024}.
Reconfigurable intelligent surfaces (RIS) have emerged as a transformative technology for next-generation mobile wireless networks, offering the ability to dynamically create a favorable propagation environment \cite{ElMossallamy TCCN 2020}-\cite{Wu Qingqing Cmag 2020}. This capability is expected to overcome the limitations of direct-path sensing, enabling advanced multi-perspective sensing that enhances target detection reliability and tracking accuracy.

Motivated by the promising applications of RIS in wireless communication systems, researchers have recently explored the integration of RIS with ISAC technology \cite{Liu Rang WCOM 2023}, \cite{Meng Kaitao WCOM 2024}. In \cite{Luo Honghao TVT 2023}, a joint optimization framework is proposed for active beamforming at the base station (BS) and passive beamforming at the RIS, aimed at maximizing the achievable communication rate while preserving radar beam pattern similarity. The study in \cite{Xu Yongqing TCOM 2024} investigates the integration of RIS in ISAC systems to enhance spatial degrees of freedom (DoF) for beamforming, with the objective of optimizing radar mutual information. In \cite{Liu Rang JSTSP 2022}, the authors address the joint optimization of transmit waveforms, receive filters, and RIS reflection coefficients in RIS-assisted ISAC systems under cluttered environments. Moreover, \cite{Yuan Fang TCOM 2024} explores the deployment of multiple RISs to create independent reflective paths, thereby improving communication and sensing coverage in areas obstructed by buildings. Extending these efforts to broadband ISAC systems, \cite{Wei Tong ICOM 2023} investigates the Doppler shift robustness of multi-RIS-assisted ISAC systems for moving target detection.

The aforementioned studies highlight the significant potential and broad application prospects of RIS in ISAC systems. However, subsequent research has revealed the multiplicative fading characteristics of RIS, which pose a major challenge for the deployment of RIS-based ISAC systems. Particularly, in sensing applications, the received signal is subject to round-trip path loss, which amplifies the limitations imposed by RIS-induced multiplicative fading compared to its impact on communication functions.

Active RIS, which integrates amplifiers with passive electromagnetic components \cite{Zhang Zijian TCOM 2023}, \cite{Long Ruizhe TWC 2021}, presents a promising solution to address these challenges. Beyond simply altering the phase of the incoming signals, active RIS can amplify the signal amplitude, effectively mitigating the adverse effects of multiplicative fading. This capability positions active RIS as a superior choice for ISAC system implementation. For instance, active RIS can establish virtual line-of-sight (LoS) links for detecting targets obstructed by buildings, thereby maximizing sensing performance in terms of signal-to-interference-plus-noise ratio (SINR) \cite{Zhu Qi TVT 2023}, \cite{Yu Zhiyuan TCOM 2024}, and Cram{\'e}r-Rao bound (CRB) \cite{Zhu Qi TWC 2024}, while ensuring the maintenance of the communication quality of service (QoS). Furthermore, studies in \cite{Zhang Yang TVT 2024} and \cite{Liu Mengyu TCCN 2024} investigate the advantages of active RIS in ISAC systems, particularly in interference management and enhancing wireless communication security. Additionally, \cite{Wang Fangzhou SPL 2022} explores the integration of active RIS with passive RIS-assisted ISAC systems, offering insights into their complementary benefits.

In the aforementioned RIS-assisted ISAC systems, RIS is primarily utilized to establish additional NLoS paths, which either facilitate target detection in obstructed environments or enhance the strength of received echo signals. In clutter-rich ISAC scenarios, beamforming and spatial filtering are commonly employed to separate targets from clutter, relying predominantly on spatial DoFs. However, prior studies have largely overlooked the potential of leveraging the unique characteristics of NLoS paths in the delay and Doppler domains, which are influenced by geometric positioning and observational perspectives.

Observing a moving target from multiple perspectives or paths, each exhibiting distinct spatial, delay, and Doppler characteristics, offers substantial advantages in distinguishing the target from clutter. This benefit arises from the diversity inherent in the spatial, delay, and Doppler domains, which can significantly enhance the performance of ISAC systems in complex urban environments. Recent studies \cite{Zuo Lei TAES 2024}-\cite{Tang Bo TSP 2024} have demonstrated that space-time adaptive processing (STAP), applied across both spatial and temporal (delay and Doppler) domains, can effectively improve clutter suppression and target detection by leveraging the unique spatial, delay, and Doppler signatures of both the target and clutter. Furthermore, the deployment of active RIS enables the adaptive generation of high-quality NLoS observation paths, facilitating multi-perspective observations that provide diversity across the spatial, delay, and Doppler domains. This highlights the critical importance of investigating the integration of STAP and RIS technologies in ISAC systems to fully exploit the advantages of multi-perspective observations and achieve superior performance in challenging environments.

Motivated by the above discussions, this paper investigates a more practical multi-active-RIS assisted ISAC system, where the BS aims to detect a moving target in the presence of static clutters. We utilize constructive interference (CI)-based symbol-level precoding (SLP) for waveform design at the transmit side and spatial-temporal adaptive processing at the receive end. Unlike conventional beamforming techniques that rely solely on spatial-domain separation to distinguish targets from clutter, the proposed approach exploits the DoFs in the spatial, delay, and Doppler domains, thereby significantly enhancing clutter suppression and target detection performance. The main contributions of this paper are summarized as follows:

\begin{itemize}
\item We develop a comprehensive signal model that characterizes the spatial, delay, and Doppler domain features of both LoS and NLoS paths facilitated by active RISs, providing a foundation for enhanced moving target sensing in multi-perspective scenarios. To fully exploit the potential of each observation path and leverage multi-perspective observation, we propose a joint design framework encompassing transmit space-time waveforms, active RIS reflection coefficients, and receive space-time filters. The objective is to maximize the radar receive SCNR after applying STAP, while simultaneously ensuring the QoS requirements for multi-user multiple-input single-output (MU-MISO) communication systems.

\item We propose an alternating iterative framework to solve the non-convex optimization problem efficiently. Specifically, the minimum variance distortionless response (MVDR) technique is employed to derive a closed-form solution for the receive filter. Subsequently, the Dinkelbach transformation is utilized to reformulate the fractional objective function into a polynomial form, facilitating its tractability. Finally, we develop an iterative algorithm that integrates the majorization-minimization (MM) method with the alternating direction method of multipliers (ADMM) to jointly optimize the transmit waveform and the reflection coefficients of the RIS.

\item We present comprehensive simulation results to validate the effectiveness of the proposed multi-perspective observation strategy in enhancing target detection performance in the presence of substantial static clutter. The impact of key system parameters, including the location and number of RISs, as well as the magnitude and direction of target velocity, is systematically evaluated. The results underscore the robustness of the proposed approach, demonstrating its potential for practical deployment in complex urban environments where static clutter presents a significant challenge.

\end{itemize}

\textit{Notation}:
Boldface lower-case and upper-case letters indicate column vectors and matrices, respectively.
$(\cdot)^T$, $(\cdot)^\ast$, $(\cdot)^H$ and $(\cdot)^{-1}$ denote the transpose, conjugate, transpose-conjugate and inverse operations, respectively.
$\mathbb{C}$ and $\mathbb{R}$ denote the sets of complex numbers and real numbers, respectively.
$| a |$ and $\|\mathbf{a}\|$ are the magnitude of a scalar $a$, and the 2-norm of a vector $\mathbf{a}$ respectively.
$\angle{a}$ is the angle of complex-valued $a$.
$\Re\{a\}$ and $\Im\{a\}$ denote the real and imaginary part of a scalar $a$, respectively.
$\otimes$ denotes the Kronecker product.
$\mathbb{E}\{\cdot\}$ denotes the expectation operation.
$\text{Tr}\{\mathbf{A}\}$ takes the trace of a matrix $\mathbf{A}$ and $\text{vec}\{\mathbf{A}\}$ vectorizes the matrix $\mathbf{A}$.
$\mathbf{I}_M$ indicates an $M\times M$ identity matrix and $\mathbf{1}_M$ denotes the $M \times 1$ vector with all entries being 1.

\section{System Model and Problem Formulation}\label{sec:system model}

\begin{figure}[!t]
\centering
\includegraphics[width = \linewidth]{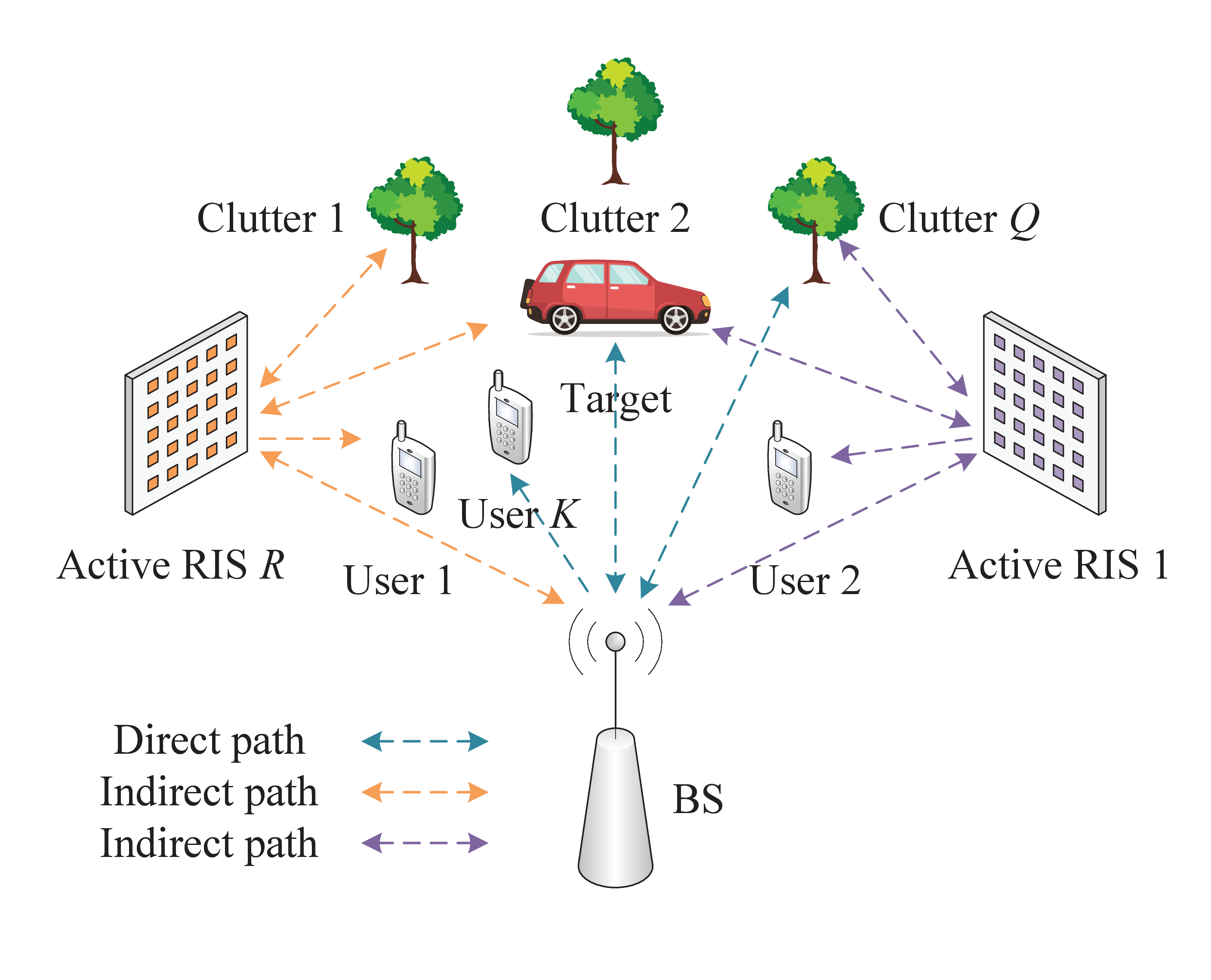}\vspace{0.0 cm}
\caption{The active RIS assisted ISAC system.}
\label{fig:system_model}
\end{figure}

This paper considers an active-RIS-empowered ISAC system, as depicted in Fig. \ref{fig:system_model}. Specifically, the dual-functional BS is equipped with $N$ transmit antennas and $N$ receive antennas. The transmit/receive antennas are arranged in a uniform linear array (ULA) with half-wavelength spacing.  When BS provides downlink communication services to $K$ single-antenna users, it  concurrently detects a moving target with the aid of $R$ active RISs. The surveillance environment encompasses a sophisticated blend of urban and rural landscapes, where the target is surrounded by $Q$ stationary clutters.
Spatially arranged active RISs, each comprising \( N_\mathrm{r} \) elements, can generate multiple high-quality controllable NLoS propagation paths, thereby providing several sensing perspectives that enhance target detection in complex and clutter-enriched environments.
To simplify the system representation and algorithm development, we define the sets $\mathcal{R}\triangleq\{1,2,\cdots, R\}$ and
$\mathcal{Q}\triangleq\{1,2,\cdots,Q\}$ to respectively represent the indices of active RISs and clutters.

Due to the large number of scatterers, the direct path is highly susceptible to significant interference, which negatively affects the target detection performance. Fortunately, by strategically deploying active RISs, controllable NLoS paths introduce additional sensing observation perspectives with distinct time delays and Doppler shifts. These multiple sensing signal propagation paths, including both NLoS and LoS paths, provide diversity of perspectives of observation, thereby facilitating the suppression of strong clutter interference. The proposed multi-perspective observation approach leverages the spatial, delay, and Doppler characteristics of each path to enhance target detection performance, achieved through joint optimization of the transmission waveform, reflection coefficients of active RISs, and receiving filters.

Specifically, we assume that the BS transmits
$M$ pulses at a constant pulse repetition frequency (PRF) during a coherent processing interval (CPI) with each pulse $\mathbf{X}_m$  divided into $L$ time slots, denoted as $\mathbf{X}_m\triangleq [\mathbf{x}_m[1], \mathbf{x}_m[2], \cdots, \mathbf{x}_m[L]] \in \mathbb{C}^{N\times L}, \forall m$.
In order to achieve both information transmission and efficient utilization of the spatial and temporal design DoFs of the transmit waveform, the dual-functional BS employs the SLP technique to generate the waveform $\mathbf{X}\triangleq [\mathbf{X}_1, \mathbf{X}_2, \cdots, \mathbf{X}_M]$ for all pulses. This approach not only satisfies the quality of service (QoS) requirements for communication users but also enables phase encoding of radar pulses in both the fast and slow time dimensions.

\subsection{Radar Model}

From the perspective of radar sensing, the BS has prior knowledge of the location of the $r$-th active RIS, $\forall r \in \mathcal{R}$, as the RISs are pre-deployed in the environment. The target, moving at a constant velocity, is surrounded by
$Q$ static clutters, generated by objectives such as buildings, trees, and streetlights, typical of a complex urban environment.
Given the persistent presence of these obstructions, a dynamic environmental database can be developed to obtain prior clutter information, leveraging techniques such as geographic information systems or ray tracing.

With the deployment of active RISs, the received echo signal at the BS is a superposition of multiple path components, each experiencing distinct delays and Doppler shifts. These unique path attributes are crucial to the proposed multi-perspective observation technique. Consequently, we develop a comprehensive received signal model to capture these characteristics.
For the transmitted signal $\mathbf{x}_m[l]$ at the $m$-th pulse and the $l$-th time slot, the snapshot signal received by the BS can be represented as:
\begin{equation}\label{eq:echo shapshot}
\mathbf{y}_m[l]=\mathbf{y}_{\mathrm{t},m}[l]+\mathbf{y}_{\mathrm{c},m}[l]+\mathbf{z}_{m}[l],
\end{equation}
where $\mathbf{y}_{\mathrm{t},m}[l]$, $\mathbf{y}_{\mathrm{c},m}[l]$, and $\mathbf{z}_{m}[l]$ denote the signal components corresponding to target echo signals, clutter echo signals, and additive white Gaussian noise (AWGN), respectively.

The proposed multi-perspective approach ensures that the target echo incorporates signals from both the direct LoS path and the indirect NLoS paths enabled by active RISs. As shown in Fig. \ref{fig:system_model}, the echo from the indirect paths to the target consists of three distinct components: BS-target-active RIS-BS, BS-active RIS-target-BS, and BS-active RIS-target-active RIS-BS.
We define a set $\mathcal{I} \triangleq \{1,2,3\}$ to index each NLoS path introduced by the active RISs.
Therefore, the target echo signal $\mathbf{y}_{\mathrm{t},m}[l]$, which encompasses all components of the path, can be explicitly expressed as
\begin{equation}\label{eq:target snapshot m l}
\begin{aligned}
\mathbf{y}_{\mathrm{t},m}[l]&=e^{\jmath(m-1)f_{\mathrm{d},0}T}\mathbf{H}_{\mathrm{t},0}\mathbf{x}_m[l-\tau_{\mathrm{t},0}]\\
&+\sum_{r \in \mathcal{R}}\sum_{i \in \mathcal{I}}e^{\jmath(m-1)f_{\mathrm{d},r,i}T}\mathbf{H}_{\mathrm{t},r,i}(\bm{\phi}_r)\mathbf{x}_m[l-\tau_{\mathrm{t},r,i}].
\end{aligned}
\end{equation}
The target's direct path channel $\mathbf{H}_{\mathrm{t},0}$ and indirect path channels $\mathbf{H}_{\mathrm{t},r,i}(\bm{\phi}_r)$ via the $r$-th active RIS are defined as follows to simplify the notation:
\begin{subequations}\label{eq:define Htri}
\begin{align}
\mathbf{H}_{\mathrm{t},0}&\triangleq \alpha_{\mathrm{t},0} \mathbf{a}(\theta_\mathrm{t})\mathbf{a}^T(\theta_\mathrm{t}),\\
\mathbf{H}_{\mathrm{t},r,1}(\bm{\phi}_r)&\triangleq \alpha_{\mathrm{t},r,1}\mathbf{G}_r^T\bm{\Phi}_r\mathbf{b}(\theta_{\mathrm{t},r})\mathbf{a}^T(\theta_\mathrm{t}), ~\forall r ,\\
\mathbf{H}_{\mathrm{t},r,2}(\bm{\phi}_r)&\triangleq \alpha_{\mathrm{t},r,1}\mathbf{a}(\theta_\mathrm{t})\mathbf{b}^T(\theta_{\mathrm{t},r})\bm{\Phi}_r\mathbf{G}_r, ~\forall r ,\\
\mathbf{H}_{\mathrm{t},r,3}(\bm{\phi}_r)&\triangleq \alpha_{\mathrm{t},r,3} \mathbf{G}_r^T\bm{\Phi}_r\mathbf{b}(\theta_{\mathrm{t},r})\mathbf{b}^T(\theta_{\mathrm{t},r})\bm{\Phi}_r\mathbf{G}_r,~\forall r.
\end{align}
\end{subequations}
The vectors $\mathbf{a}(\theta)$ and $\mathbf{b}(\theta)$ are the transmit/receive steering vector for the BS and active RIS at angle $\theta$, respectively:
\begin{subequations}\begin{align}
\mathbf{a}(\theta) &\triangleq \big[1,e^{-\jmath \pi\sin\theta},\ldots,e^{-\jmath \pi(N-1)\sin\theta}\big]^T,\\
\mathbf{b}(\theta) &\triangleq \big[1,e^{-\jmath \pi \sin\theta},\ldots,e^{-\jmath \pi (N_\mathrm{r}-1)\sin\theta}\big]^T.
\end{align}\end{subequations}
The matrix $\mathbf{G}_r \in \mathbb{C}^{N_\mathrm{r}\times N}$ denotes the channel between the BS and the $r$-th active RIS, and $\bm{\Phi}_r\in \mathbb{C}^{N_\mathrm{r}\times N_\mathrm{r}}$ denotes the active RIS reflection coefficient matrix with
$\bm{\Phi}_r\triangleq \mathrm{diag}\{\bm{\phi}_r\}$, where $\bm{\phi}_r\triangleq [\phi_{r,1},\cdots, \phi_{r,N_\mathrm{r}}]^T$ is the reflection coefficient vector.
To effectively mitigate the severe path loss arising from both multiplicative fading and round-trip echo signals, active RISs embed amplifiers within traditionally passive electromagnetic components. This integration allows the reflection coefficients to be tuned in both phase and amplitude simultaneously, thereby relaxing the constant-modulus constraint of $\bm{\Phi}_r$.
Moreover, in the compound channel model \eqref{eq:define Htri}, the scalars $\alpha_{\mathrm{t},0}$ and $\alpha_{\mathrm{t},r,i}, ~\forall r, ~\forall i$ represent the path-loss of the target's echo on the direct path and each indirect path, respectively. The coefficient $f_{\mathrm{d},0}$ denotes the Doppler frequency observed from the direct path, while the coefficient $f_{\mathrm{d},r,i}$ represents the Doppler frequency observed from the $i$-th indirect path via the $r$-th active RIS. The method for calculating the Doppler frequency of each path is detailed in \cite{Wei Tong ICOM 2023}.
The coefficients $\tau_{\mathrm{t},0}$ and $\tau_{\mathrm{t},r,i}$ respectively stand for the discretized time delay for the direct path and the indirect paths related to the $r$-th active RIS in the fast-time domain, which can be calculated by the formula in \cite{Xie Zhuang TSP 2024}.
It is noteworthy that each propagation path can induce distinct time delays and Doppler shifts, both of which are influenced by the underlying geographical conditions and geometric relationships. This inherent diversity represents a key advantage of the proposed multi-perspective observation strategy.

Similarly to the target echo signal (\ref{eq:target snapshot m l}), the clutter echo consists of multiple components from both direct and indirect paths. However, unlike the target, the static nature of the clutters ensures that no Doppler shift is induced. Hence, the clutter signal in the $l$-th time slot during the $m$-th pulse can be expressed as
\begin{equation}
\mathbf{y}_\mathrm{c}[l]=\! \sum_{q \in \mathcal{Q}}(\mathbf{H}_{q,0}\mathbf{x}_m[l-\tau_{q,0}]+\!\sum_{r \in \mathcal{R}}\sum_{i \in \mathcal{I}}\!\mathbf{H}_{q,r,i}(\bm{\phi}_r)\mathbf{x}_m[l-\tau_{q,r,i}]).
\end{equation}
It is assumed that prior information on the geographical location of the clutters is available at the BS. Consequently, the coefficients $\tau_{q,0}$ and $\tau_{q,r,i}$, representing the discretized time delay for the direct and indirect paths of the clutter, can be calculated by the geographical location.
Additionally, we define the equivalent channels associated with the $q$-th clutter as
\begin{subequations}\label{eq:define Hqri}
\begin{align}
\mathbf{H}_{q,0}&\triangleq \alpha_{q,0}\mathbf{a}(\theta_q)\mathbf{a}^T(\theta_q),\\
\mathbf{H}_{q,r,1}(\bm{\phi}_r)&\triangleq \alpha_{q,r,1} \mathbf{G}_r^T\bm{\Phi}_r\mathbf{b}(\theta_{q,r})\mathbf{a}^T(\theta_q),\\
\mathbf{H}_{q,r,2}(\bm{\phi}_r)&\triangleq \alpha_{q,r,2} \mathbf{a}(\theta_q)\mathbf{b}^T(\theta_{q,r})\bm{\Phi}_r \mathbf{G}_r,\\
\mathbf{H}_{q,r,3}(\bm{\phi}_r) &\triangleq \alpha_{q,r,3} \mathbf{G}_r^T\bm{\Phi}_r\mathbf{b}(\theta_{q,r})\mathbf{b}^T(\theta_{q,r})\bm{\Phi}_r\mathbf{G}_r,
\end{align}
\end{subequations}
where $\alpha_{q,0}$ represents the path-loss of the $q$-th clutter's direct path, while  $\alpha_{q,r,i}$ denotes the $q$-th clutter's indirect path associated with the $r$-th RIS.
In addition, the vector $\mathbf{z}_m[l]$ in  (\ref{eq:echo shapshot}) is the AWGN and
$\mathbf{z}_m[l]\sim \mathcal{CN}(\mathbf{0},\sigma_\mathrm{z}^2\mathbf{I}_N)$.

To leverage the benefits of the proposed multi-perspective observation strategy for moving target detection, joint space-time processing is applied to the received signals. This involves stacking and stretching operations on the received snapshot signals corresponding to various reception pulses and time slots. We begin by stacking the received signals from the
$m$-th pulse, which are represented as a matrix $\mathbf{Y}_m \triangleq \big[\mathbf{y}_m[1], \cdots, \mathbf{y}_m[l], \cdots, \mathbf{y}_m[P]\big] \in \mathbb{C}^{N\times P}$.
It is proposed that $P\geq \mathrm{max}\{\tau_{\mathrm{t},r,i}, \forall r, \forall i\}$ snapshots are received to effectively capture the target echo signals from each path, as discussed in \cite{Xie Zhuang TSP 2024}.
Thus, the received signal from the $m$-th pulse can be expressed as
\begin{equation}\label{eq: define Ym}
\mathbf{Y}_m\triangleq \mathbf{Y}_{\mathrm{t},m}+\mathbf{Y}_{\mathrm{c},m}+\mathbf{Z}_m.
\end{equation}
The matrices $\mathbf{Y}_{\mathrm{t},m}$ and $\mathbf{Y}_{\mathrm{c},m}$ represent the echo signals of the target and clutters in different range bins within the same pulse, respectively.
For the target, we consider that the Doppler shift within the same pulse remains constant.
Thus, the matrices $\mathbf{Y}_{\mathrm{t},m}$ and $\mathbf{Y}_{\mathrm{c},m}$ can be expressed as
\begin{equation}\label{eq:target echo from the same pulse}
\begin{aligned}
\mathbf{Y}_{\mathrm{t},m}&=e^{\jmath(m-1)f_{\mathrm{d},0}T}\mathbf{H}_{\mathrm{t},0}\mathbf{X}_m\mathbf{J}_{\mathrm{t},0}\\
&\quad~ +\sum_{r \in \mathcal{R}}\sum_{i \in \mathcal{I}}e^{\jmath(m-1)f_{\mathrm{d},r,i}T}\mathbf{H}_{\mathrm{t},r,i}(\bm{\phi}_r)\mathbf{X}_m\mathbf{J}_{\mathrm{t},r,i},
\end{aligned}
\end{equation}
\begin{equation}\label{eq:clutter echo from the same pulse}
\mathbf{Y}_{\mathrm{c},m}=\sum_{q \in \mathcal{Q}} (\mathbf{H}_{q,0}\mathbf{X}_m \mathbf{J}_{q,0}+\sum_{r \in \mathcal{R}}\sum_{i \in \mathcal{I}}\mathbf{H}_{q,r,i}(\bm{\phi}_r)\mathbf{X}_m \mathbf{J}_{q,r,i}).
\end{equation}
The shift matrices $\mathbf{J}_{\mathrm{t},r,i}\in \mathbb{R}^{L\times P}$ and $\mathbf{J}_{q,r,i}\in \mathbb{R}^{L\times P}$ are introduced to indicate the relative delay between the direct path and the $i$-th indirect path of the $r$-th RIS for the target and $q$-th clutter, respectively.
Taking $\mathbf{J}_{\mathrm{t},r,i}$ as an example, it is defined as follows
\begin{equation}
\mathbf{J}_{\mathrm{t},r,i}(m, n)= \begin{cases}1, & m-n+\tau_{\mathrm{t},r,i}-\tau_{\mathrm{t},0}=0 , \\ 0, & \text { otherwise}.\end{cases}
\end{equation}

In order to exploit the DoFs in the Doppler dimension, we stack and vectorize echo signal of all $M$ pulses  by defining $\mathbf{y}\triangleq \mathrm{vec}\{[\mathbf{Y}_{1}, \mathbf{Y}_{2},\cdots, \mathbf{Y}_{M}]\} \in \mathbb{C}^{NMP\times 1}$. Thus the
vectorized received signal $\mathbf{y}$ can be expressed as
\begin{align}\label{eq:vector y}
\mathbf{y}=\mathbf{y}_\mathrm{t}+\mathbf{y}_\mathrm{c}+\mathbf{z}.
\end{align}
By utilizing the properties of Kronecker product \cite{Horn 1990}, the target echo signal $\mathbf{y}_\mathrm{t}$ in  (\ref{eq:vector y}) can be equivalently expressed as
\begin{equation}\label{eq:define yt}
\mathbf{y}_\mathrm{t}\triangleq \big(\widetilde{\mathbf{H}}_\mathrm{dir}+\widetilde{\mathbf{H}}_\mathrm{ind}(\bm{\phi})\big)\mathbf{x},
\end{equation}
where $\mathbf{x} \triangleq \mathrm{vec}\{\mathbf{X}\} \in \mathbb{C}^{NML\times 1}$ is the vectorized transmit waveforms. For conciseness,  $\widetilde{\mathbf{H}}_\mathrm{dir}$ and $\widetilde{\mathbf{H}}_\mathrm{ind}$ denote direct and indirect channel matrices, which are defined as
\begin{subequations}
\begin{align}
\widetilde{\mathbf{H}}_\mathrm{dir} &\triangleq \mathbf{D}(f_{\mathrm{d},0})\otimes \mathbf{J}_{\mathrm{t},0}^T \otimes \mathbf{H}_{\mathrm{t},0},\\
\widetilde{\mathbf{H}}_\mathrm{ind}(\bm{\phi}) &\triangleq \sum_{r \in \mathcal{R}} \sum_{i \in \mathcal{I}} \mathbf{D}(f_{\mathrm{d},r,i})\otimes \mathbf{J}_{\mathrm{t},r,i}^T\otimes \mathbf{H}_{\mathrm{t},r,i}(\bm{\phi}_r).\label{eq:define Hdir Hind}
\end{align}
\end{subequations}
The matrix $\mathbf{D}(f_{\mathrm{d}})$ is the Doppler shift matrix, with the definition as
\begin{equation}
\mathbf{D}(f_{\mathrm{d}})\triangleq \mathrm{diag}\{1, e^{\jmath f_{\mathrm{d}}T}, \cdots, e^{\jmath (m-1)f_{\mathrm{d}}T}\}.
\end{equation}
We also define $\bm{\phi}\triangleq [\bm{\phi}_1^T, \bm{\phi}_2^T,\ldots,\bm{\phi}_R^T]^T$ for notational simplicity.
Compared to the target echo signal, the clutter echo signal $\mathbf{y}_\mathrm{c}$ also has a similar structure, except that the Doppler shift matrices for the clutters are the identity matrix.
Therefore, $\mathbf{y}_\mathrm{c}$ in  (\ref{eq:vector y}) can be expressed as
\begin{align}
\mathbf{y}_\mathrm{c}\triangleq \big(\widetilde{\mathbf{H}}_{\mathrm{dir},\mathrm{c}}+\widetilde{\mathbf{H}}_{\mathrm{ind},\mathrm{c}}(\bm{\phi})\big)\mathbf{x},
\end{align}
where we define
\begin{align}
\widetilde{\mathbf{H}}_{\mathrm{dir},\mathrm{c}}& \triangleq \sum_{q \in \mathcal{Q}} \mathbf{I}_M \otimes \mathbf{J}^T_{q,0} \otimes \mathbf{H}_{q,0},\\
\widetilde{\mathbf{H}}_{\mathrm{ind},\mathrm{c}}(\bm{\phi})& \triangleq \sum_{q \in \mathcal{Q}} \sum_{r \in \mathcal{R}}\sum_{i \in \mathcal{I}}(\mathbf{I}_M \otimes \mathbf{J}^T_{q,r,i} \otimes \mathbf{H}_{q,r,i}(\bm{\phi}_r)).
\end{align}
Besides, the vector $\mathbf{z}\triangleq \mathrm{vec}\{[\mathbf{Z}_1, \mathbf{Z}_2, \cdots, \mathbf{Z}_M]\}$ denotes the AWGN with $\mathbf{z}\thicksim \mathcal{CN}(\mathbf{0},\sigma_\mathrm{r}^2\mathbf{I}_{NMP})$.

After stacking and vectorizing the received signals, they are processed by a linear spatial-temporal receive filter $\mathbf{w}\in\mathbb{C}^{MNP\times 1}$.
Then, the target detection task can be formulated as a binary hypothesis testing problem as
\begin{equation}
\left\{\begin{array}{l}
\mathcal{H}_0: \mathbf{w}^H\mathbf{y}=\mathbf{w}^H(\mathbf{y}_{\mathrm{c}}+\mathbf{z}), \\
\mathcal{H}_1: \mathbf{w}^H\mathbf{y}=\mathbf{w}^H(\mathbf{y}_\mathrm{t}+\mathbf{y}_{\mathrm{c}}+\mathbf{z}).
\end{array}\right.
\end{equation}
The null hypothesis $\mathcal{H}_0$ and the alternative hypothesis $\mathcal{H}_1$ represent the two states of the target being absent/present, respectively.
According to the Neyman-Pearson criterion, in the presence of Gaussian noise, the probability of target detection increases monotonically with the output SCNR of the spatial-temporal receive filter, given that the false alarm probability is constrained not to exceed a certain threshold.
Therefore, in this paper, we utilize the output SCNR of radar filter  as our sensing performance metric, which can be calculated as
\begin{equation}\label{eq:gamma}
\mathrm{SCNR}_\mathrm{r}=\frac{|\mathbf{w}^H (\widetilde{\mathbf{H}}_\mathrm{dir}+\widetilde{\mathbf{H}}_\mathrm{ind}(\bm{\phi}))\mathbf{x}|^2}{|\mathbf{w}^H(\widetilde{\mathbf{H}}_{\mathrm{dir},\mathrm{c}}+\widetilde{\mathbf{H}}_{\mathrm{ind},\mathrm{c}}(\bm{\phi}))\mathbf{x}|^2+\sigma^2_\mathrm{r}\mathbf{w}^H\mathbf{w}}.
\end{equation}

\subsection{Communication Model}
In addition to its sensing capabilities, the dual-functional BS provides downlink communication services to $K$ single-antenna users, thereby achieving enhanced spectrum utilization and hardware efficiency during the sensing process.
Specifically, the BS realizes the transmission of information symbols by modulating the transmitted waveforms over both fast and slow time scales using SLP technique.
During the $m$-th transmission pulse and within the $l$-th time slot, the symbol vector intended to the $K$ users is denoted as $\mathbf{s}_{m}[l] \triangleq [s_{m,1}[l],\ldots,s_{m,K}[l]]^T$, where each symbol is assumed to be independently chosen from an $\Omega$-phase shift keying (PSK) constellation.
The relationship between the transmitted symbol vector $\mathbf{s}_{m}[l]$ and the waveform $\mathbf{x}_{m}[l]$ emitted from $N$ antennas is a complex nonlinear mapping,
which will be discussed in detail in the following paragraph.
The corresponding received signal in the $l$-th time slot of the $m$-th pulse at the $k$-th user can be expressed as
\begin{equation}
y_{m,k}[l]=\big(\mathbf{h}^T_{\mathrm{d},k}+\sum_{r \in \mathcal{R}}\mathbf{h}^T_{r,k}\bm{\Phi}_r\mathbf{G}_r\big)\mathbf{x}_m[l]+n_{m,k}[l],
\end{equation}
where $\mathbf{h}_{\mathrm{d},k} \in \mathbb{C}^{N\times 1}$ represents the LoS channel from the BS to the $k$-th user, and $\mathbf{h}_{r,k} \in \mathbb{C}^{N_\mathrm{r}\times 1}$
denotes the channel from the $r$-th RIS to the $k$-th user.
The scalar $n_{m,k}[l]\sim\mathcal{CN}(0,\sigma^2_k)$ represents the AWGN at the $k$-th user.

\begin{figure}[!t]
\centering
\includegraphics[width = 2.8 in]{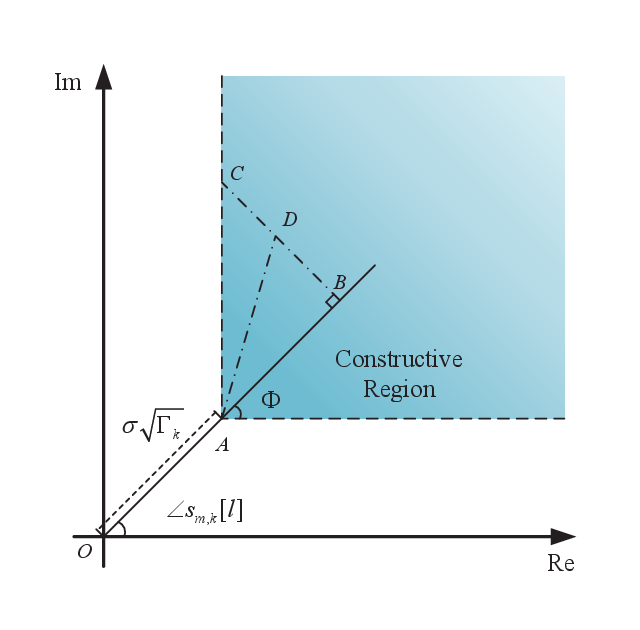}\vspace{-0.2 cm}
\caption{CI-based SLP for QPSK constellation.}\label{fig:CR}
\end{figure}

We use the constructive interference (CI)-based SLP approach as an example, as illustrated in Fig. \ref{fig:CR}, to demonstrate the method for ensuring communication QoS.
In Fig. \ref{fig:CR}, the desired information symbol for the $k$-th user is $(1/\sqrt{2}, \jmath/\sqrt{2})$, and the blue area represents its constructive region.
Let point $D$ denote the received noise-free signal for the $k$-th user $\widetilde{y}_{m,k}[l] = (\mathbf{h}^T_{\mathrm{d},k}+\sum_{r \in \mathcal{R}}\mathbf{h}^T_{r,k}\bm{\Phi}_r\mathbf{G}_r)\mathbf{x}_m[l]$.
When the QoS requirement for the $k$-th users is $\Gamma_k$, if the multi-user interference is completely eliminated, the received noise-free signal should be located at point $A$, which satisfies $\widetilde{y}_{m,k}[l] = \sigma_k\sqrt{\Gamma_{k}}s_{m,k}[l]$.
Rather than suppressing interference, the CI-based SLP approach exploits beneficial multi-user interference (MUI) to push the received noise-free signal further into the constructive region. Specifically, after SLP processing, the noise-free signal $\widetilde{y}_{m,k}[l]$ must fall within the constructive region, which can be geometrically expressed as $|\overrightarrow{BC}| > |\overrightarrow{BD}|$.
Due to space limitations, we refer readers to \cite{MA ICST 2018}–\cite{Liu TWC 2021} for further derivation details.
The QoS constraints that guarantee that the noise-free received signal $\widetilde{y}_{m,k}[l]$ lies in the constructive region can be expressed as
\begin{align}\label{eq:safety margin}
& \Re\big\{(\mathbf{h}^T_{\mathrm{d},k}\!+\!\sum_{r \in \mathcal{R}}\mathbf{h}^T_{r,k}\bm{\Phi}_r\mathbf{G}_r) \mathbf{x}_m[l] e^{-\jmath \angle s_{m,k}[l]}\!-\!\sigma_{k} \sqrt{\Gamma_{k}}\big\} \tan \Phi \\ \nonumber
& -\big|\Im\big\{(\mathbf{h}^T_{\mathrm{d},k}\!+\!\sum_{r \in \mathcal{R}}\mathbf{h}^T_{r,k}\bm{\Phi}_r\mathbf{G}_r) \mathbf{x}_m[l] e^{-\jmath \angle s_{m,k}[l]}\big\}\big| \geq 0,
\end{align}
where $\Phi = \pi/\Omega$.
A compact form for the communication QoS requirement (\ref{eq:safety margin}) can be written as
\begin{equation}\label{eq:communication constraint}
\Re\big\{\widetilde{\mathbf{h}}_i^T(\bm{\phi})\mathbf{x}\}\geq \gamma_i,~~\forall i = 1, \ldots, 2KML.
\end{equation}
The vector $\widetilde{\mathbf{h}}_i(\bm{\phi})$ in (\ref{eq:communication constraint}) represents the overall equivalent channel for the transmit waveform $\mathbf{x}$, which defined by
\begin{subequations}
\begin{align}\label{eq:define hitilde phi}
\widetilde{\mathbf{h}}_{(2k-1)ML+j}^T(\bm{\phi})&\triangleq \xi_{m,k,1}\mathbf{e}_{j,ML}^T\otimes\mathbf{h}^T_{k}(\bm{\phi}), \\
\widetilde{\mathbf{h}}_{(2k-2)ML+j}^T(\bm{\phi})&\triangleq \xi_{m,k,2}\mathbf{e}_{j,ML}^T\otimes\mathbf{h}^T_{k}(\bm{\phi}).
\end{align}
\end{subequations}
For conciseness, we define
\begin{subequations}\label{eq:define hk xi}
\begin{align}
\mathbf{h}^T_{k}(\bm{\phi})&\triangleq \mathbf{h}^T_{\mathrm{d},k}+\sum_{r \in \mathcal{R}}\bm{\phi}_r^T\mathrm{diag}\{\mathbf{h}^T_{r,k}\}\mathbf{G}_r,\\
\xi_{m,k,1}&\triangleq e^{-\jmath \angle s_{m,k}[l]}(\sin \Phi + e^{- \jmath \frac{\pi}{2}} \cos \Phi), \\
\xi_{m,k,2}&\triangleq e^{-\jmath \angle s_{m,k}[l]}(\sin \Phi - e^{- \jmath \frac{\pi}{2}} \cos \Phi),
\end{align}
\end{subequations}
and the vector $\mathbf{e}_{j,ML}\in\mathbb{R}^{ML}$ indicates the $j$-th column of an $ML\times ML$ identity matrix.
In addition, the parameters $\gamma_i, \forall i$ in (\ref{eq:communication constraint}) is defined by
$\gamma_{2(k-1)ML+j}\triangleq \sigma_k \sqrt{\Gamma_k}\sin{\Phi}$.

\subsection{Problem Formulation}
This paper aims to jointly design the transmit waveform $\mathbf{x}$ and the receive filter $\mathbf{w}$ to maximize the radar output SCNR (\ref{eq:gamma}), while ensuring the communication QoS requirements (\ref{eq:communication constraint}). In addition, the radar waveform must account for constraints imposed by the hardware infrastructure. A common constraint is the constant modulus waveform, which is used to avoid nonlinear distortion in amplifiers, and can be expressed as
\begin{equation}\label{eq:CE constraint}
|x_j| = \sqrt{P_\mathrm{BS}/(NML)},~~\forall j = 1,\ldots,NML.
\end{equation}
Meanwhile, the reflection coefficients of the active RISs
cannot exceed the amplifiers' maximum amplitude.
Therefore, the optimization problem is formulated as
\begin{subequations}\label{eq:original problem}
\begin{align}
&\max_{\mathbf{x},\bm{\phi}, \mathbf{w}}~~~ \frac{|\mathbf{w}^H (\widetilde{\mathbf{H}}_\mathrm{dir}+\widetilde{\mathbf{H}}_\mathrm{ind}(\bm{\phi}))\mathbf{x}|^2}{|\mathbf{w}^H(\widetilde{\mathbf{H}}_{\mathrm{dir},\mathrm{c}}+\widetilde{\mathbf{H}}_{\mathrm{ind},\mathrm{c}}(\bm{\phi}))\mathbf{x}|^2+\sigma^2_\mathrm{r}\mathbf{w}^H\mathbf{w}}\\
&~\mathrm{s.t.}~~~~~\Re \{\widetilde{\mathbf{h}}_i^T(\bm{\phi})\mathbf{x}\} \geq \gamma_i, ~\forall i, \\
&~~~~~~~~~~  |x_j|=\sqrt{P_\mathrm{BS}/(NML)},~\forall j,\\
&~~~~~~~~~~  |\phi_{n}| \leq a_\mathrm{max},~\forall n,
\end{align}
\end{subequations}
where $P_\text{BS}$ is the total power budget at the BS and $a_\text{max}$ is the maximum amplitude of each reflection coefficient.
We observe that (\ref{eq:original problem}) is a complex non-convex optimization problem due to the non-convex objective function (\ref{eq:original problem}), the non-convex waveform constraints (\ref{eq:original problem}c), and the coupling among the variables $\mathbf{w}$, $\mathbf{x}$ and $\bm{\phi}$, which prevent a direct closed-form solution. To address these challenges, we decompose the waveform, active RIS reflection, and receive filter design problems into tractable sub-problems, and then develop efficient algorithms to iteratively solve them.

\section{Joint Waveform, RIS Reflection and Receive Filter Design}\label{sec:joint design}
In this section, we propose an efficient alternative optimization algorithm for the joint transmit waveform, active RIS reflection and receive filter design.
The original problem is first decomposed into three sub-problems and efficient algorithms are developed to iteratively solve them.
\subsection{Receive Filter Design and FP Transformation}
With the fixed transmit waveform $\mathbf{x}$, active RIS reflection vector $\bm{\phi}$, the optimization problem with respect to $\mathbf{w}$ becomes a minimum variance distortionless response (MVDR) problem, which can be formulated as
\begin{subequations}\label{eq:MVDR for w }
\begin{align}
\min_{\mathbf{w}}&\ \ \mathbf{w}^H(\mathbf{R}_\mathrm{c}+\sigma_\mathrm{r}^2\mathbf{I}_{NMP})\mathbf{w}\\
\text{s.t.}&\ \  \mathbf{w}^H(\widetilde{\mathbf{H}}_{\mathrm{dir}}+\widetilde{\mathbf{H}}_{\mathrm{ind}}(\bm{\phi}))\mathbf{x}=1.
\end{align}
\end{subequations}
The matrix $\mathbf{R}_\mathrm{c}$ represents the covariance matrix of the clutters, which can be expressed as
\begin{equation}
\mathbf{R}_\mathrm{c}\triangleq (\widetilde{\mathbf{H}}_\mathrm{dir,c}+\widetilde{\mathbf{H}}_\mathrm{ind,c}(\bm{\phi}))\mathbf{x}\mathbf{x}^H(\widetilde{\mathbf{H}}^H_\mathrm{dir,c}+\widetilde{\mathbf{H}}^H_\mathrm{ind,c}(\bm{\phi})).
\end{equation}
The closed-form optimal solution $\mathbf{w}^\star$ for the problem (\ref{eq:MVDR for w }) can be easily obtained as
\begin{equation}\label{eq:optimal w}
\mathbf{w}^\star = \frac{(\mathbf{R}_\mathrm{c}+\sigma_\mathrm{r}^2\mathbf{I}_{NMP})^{-1}(\widetilde{\mathbf{H}}_{\mathrm{dir}}+\widetilde{\mathbf{H}}_{\mathrm{ind}}(\bm{\phi}))\mathbf{x}}{\|(\mathbf{R}_\mathrm{c}+\sigma_\mathrm{r}^2\mathbf{I}_{NMP})^{-1}(\widetilde{\mathbf{H}}_{\mathrm{dir}}+\widetilde{\mathbf{H}}_{\mathrm{ind}}(\bm{\phi}))\mathbf{x}\|^2}.
\end{equation}

Given the receive filter $\mathbf{w}$, the waveform and active RIS reflection design is a non-convex problem due to the complex fractional term in the objective function (\ref{eq:original problem}a).
In order to tackle this issue, we utilize the Dinkelbach's transformation and introduce an auxiliary $\eta \in \mathbb{R}$ to convert the fractional term into a polynomial
expression.
Then, the waveform and active RIS reflection design can be converted as
\begin{subequations}\label{eq:FP transformation}
\begin{align}
\max_{\mathbf{x},\bm{\phi},\eta}&\ \  |\mathbf{w}^H(\widetilde{\mathbf{H}}_{\mathrm{dir}}+\widetilde{\mathbf{H}}_{\mathrm{ind}}(\bm{\phi}))\mathbf{x}|^2-\eta\sigma_\mathrm{r}^2\mathbf{w}^H\mathbf{w} \\ \nonumber
&\quad-\eta |\mathbf{w}^H(\widetilde{\mathbf{H}}_{\mathrm{dir},\mathrm{c}}+\widetilde{\mathbf{H}}_{\mathrm{ind},\mathrm{c}}(\bm{\phi}))\mathbf{x}|^2\\
\mathrm{s.t.}&\ \ \Re \{\widetilde{\mathbf{h}}_i^T(\bm{\phi})\mathbf{x}\} \geq \gamma_i,~\forall i, \\
&\ \  |x_j|=\sqrt{P_\mathrm{BS}/(NML)},~\forall j,\\
&\ \ |\phi_{n}| \leq a_\mathrm{max},~\forall n.
\end{align}
\end{subequations}
This problem can be solved by alternatively updating the auxiliary variable $\eta$, the waveform vector $\mathbf{x}$ and the active RIS reflection $\bm{\phi}$.
Given $\mathbf{x}$ and $\bm{\phi}$, the optimal solution to $\eta$ is
\begin{equation}\label{eq: update auxiliary variable eta}
\eta^\star= \frac{|\mathbf{w}^H (\widetilde{\mathbf{H}}_{\mathrm{dir}}+\widetilde{\mathbf{H}}_{\mathrm{ind}}(\bm{\phi}))\mathbf{x}|^2}{\mathbf{w}^H\mathbf{R}_\mathrm{c}\mathbf{w}+\sigma_\mathrm{r}^2\mathbf{w}^H\mathbf{w}}.
\end{equation}

\subsection{Waveform Design}
With fixed received filter $\mathbf{w}$ and active RIS reflection vector $\bm{\phi}$, the objective function of (\ref{eq:FP transformation}a) with respect to $\mathbf{x}$ can be expressed as
\begin{equation}
f_1(\mathbf{x})= \mathbf{x}^H \mathbf{A}_\mathrm{c} \mathbf{x}-\mathbf{x}^H \mathbf{A}\mathbf{x}+c_1,
\end{equation}
where we define
\begin{align}
\mathbf{A} &\triangleq (\widetilde{\mathbf{H}}^H_{\mathrm{dir}}\!+\!\widetilde{\mathbf{H}}^H_{\mathrm{ind}}(\bm{\phi}))\mathbf{w}\mathbf{w}^H (\widetilde{\mathbf{H}}_{\mathrm{dir}}\!+\!\widetilde{\mathbf{H}}_{\mathrm{ind}}(\bm{\phi})),\\ \nonumber
\mathbf{A}_\mathrm{c} &\triangleq \eta (\widetilde{\mathbf{H}}^H_{\mathrm{dir},\mathrm{c}}\!+\!\widetilde{\mathbf{H}}^H_{\mathrm{ind},\mathrm{c}}(\bm{\phi}))\mathbf{w}\mathbf{w}^H (\widetilde{\mathbf{H}}_{\mathrm{dir},\mathrm{c}}\!+\!\widetilde{\mathbf{H}}_{\mathrm{ind},\mathrm{c}}(\bm{\phi})),\\ \nonumber
c_1 &\triangleq \eta \sigma^2_\mathrm{r}\mathbf{w}^H\mathbf{w}.
\end{align}
Then, the optimization with respect to $\mathbf{x}$ is reformulated as
\begin{subequations}\label{eq:problem for x}
\begin{align}
\min_\mathbf{x} &\ \ f_1(\mathbf{x})\\
\text{s.t.}& \ \ \Re\{\widetilde{\mathbf{h}}_i^T(\bm{\phi})\mathbf{x}\}\geq \gamma_i,~\forall i,\\
& \ \ |x_j|= \sqrt{P_\mathrm{BS}/(NML)},~\forall j.
\end{align}
\end{subequations}
We can observe that $f_1(\mathbf{x})$ is a complicated non-convex quadratic function, and $\mathbf{x}$ has a constant modulus in constraint (\ref{eq:problem for x}c), which hinder the solution to problem (\ref{eq:problem for x}).

In order to solve the non-convex objective function, we propose to use the majorization-minimization (MM) method to relax the non-convex terms $-\mathbf{x}^H\mathbf{A}\mathbf{x}$.
Specially, in the $t$-th iteration, we construct a linear upper bound as the surrogate function:
\begin{equation}
-\mathbf{x}^H\mathbf{A}\mathbf{x}\leq -2\Re\{\mathbf{x}^H\mathbf{A}\mathbf{x}^{(t)}\}+c_3,\\
\end{equation}
where $\mathbf{x}^{(t)}$ is the solution obtained in the previous iteration and  we define $c_3\triangleq (\mathbf{x}^{(t)})^H\mathbf{A}\mathbf{x}^{(t)}$.

Then, to address the non-convex constraint (\ref{eq:problem for x}c), we introduce an auxiliary variable $\bm{\psi}\triangleq [\psi_1, \psi_2, \cdots, \psi_{NML}]^T$ to decouple it and convert the problem (\ref{eq:problem for x}) to
\begin{subequations}\label{eq:x and psi problem}
\begin{align}
\min_{\mathbf{x},\bm{\psi}}&\ \ \mathbf{x}^H\mathbf{A}_\mathrm{c}\mathbf{x}-2\Re\{\mathbf{x}^H\mathbf{A}\mathbf{x}^{(t)}\}\\
\text{s.t.}&\ \ \Re\{\widetilde{\mathbf{h}}_i^T(\bm{\phi})\mathbf{x}\}\geq \gamma_i, ~\forall i,\\
&\ \ |x_j|\leq \sqrt{P_\mathrm{BS}/(NML)},~\forall j,\\
&\ \ \mathbf{x}=\bm{\psi},\\
&\ \ |\psi_j|= \sqrt{P_\mathrm{BS}/(NML)},~\forall j.
\end{align}
\end{subequations}
Problem (\ref{eq:x and psi problem}) is converted into a  separable form and we can solve it in an ADMM manner by maximizing the augmented Lagrangian (AL) function.
In specific, the AL function of (\ref{eq:x and psi problem}) is constructed as
\begin{equation}\label{eq:AL function}
\mathcal{L}(\mathbf{x},\bm{\psi}, \bm{\lambda})=\mathbf{x}^H\mathbf{A}_\mathrm{c}\mathbf{x}-2\Re\{\mathbf{x}^H\mathbf{A}\mathbf{x}^{(t)}\}+ \frac{\rho}{2}\|\mathbf{x}-\bm{\psi}+\bm{\lambda}/\rho\|^2,
\end{equation}
where $\bm{\lambda}$ is the dual variable, and $\rho > 0$ is the penalty parameter.
Then, we develop a variable update framework to maximize the AL function by alternately optimizing $\mathbf{x}$, $\bm{\psi}$, and $\bm{\lambda}$ as follows:

1) Update $\mathbf{x}$: With given $\bm{\psi}$ and $\bm{\lambda}$, the optimization for $\mathbf{x}$ can be reformulated as
\begin{subequations}\label{eq:ADMM update x}
\begin{align}
\min_{\mathbf{x}}&\ \ \mathcal{L}(\mathbf{x}, \bm{\psi},\bm{\lambda})\\
\text{s.t.}&\ \ \Re\{\widetilde{\mathbf{h}}_i^T(\bm{\phi})\mathbf{x}\}\geq \gamma_i,~\forall i,\\
&\ \ |x_j|\leq\sqrt{P_\mathrm{BS}/(NML)},~\forall j.
\end{align}
\end{subequations}
It is obvious that this is a convex problem with respect to $\mathbf{x}$, whose optimal solution $\mathbf{x}^\star$ can be obtained by CVX.

2) Update $\bm{\psi}$: After updating $\mathbf{x}$, the optimization for $\bm{\psi}$ with fixed $\mathbf{x}$ and $\bm{\lambda}$ can be expressed as
\begin{subequations}\label{eq:ADMM update y}
\begin{align}
\min_{\bm{\psi}}&\ \ \frac{\rho}{2}\|\mathbf{x}-\bm{\psi}+\bm{\lambda}/\rho\|^2\\
\text{s.t.}&\ \  |\psi_j|=\sqrt{P_\mathrm{BS}/(NML)},~\forall j.
\end{align}
\end{subequations}
Although (\ref{eq:ADMM update y}) is a non-convex problem with respect to $\bm{\psi}$, we notice that the value of the quadratic term of the objective
function (\ref{eq:ADMM update y}) remains constant because each entry of $\bm{\psi}$ has a constant modulus.
Thus the maximum value of the objective function (\ref{eq:ADMM update y}) can be attainted when $\bm{\psi}$ is aligned with the linear part $\rho \mathbf{x}+\bm{\lambda}$ as
\begin{equation}\label{eq: closed form for y}
\bm{\psi}= \sqrt{P_\mathrm{BS}/(NML)}e^{\jmath \angle(\rho \mathbf{x}+\bm{\lambda})}.
\end{equation}

3) Update $\bm{\lambda}$: After updating $\mathbf{x}$ and $\bm{\psi}$, the dual variable $\bm{\lambda}$ is
updated by
\begin{equation}\label{eq:ADMM update lambda}
\bm{\lambda}:= \bm{\lambda}+\rho(\mathbf{x}-\bm{\psi}).
\end{equation}

\subsection{Active RIS Reflection Design}
After solving the waveform optimization, we focus on the subproblem of optimizing the active RIS reflection coefficients $\bm{\phi}$.
Since the variable $\bm{\phi}$ is embedded in the equivalent channel matrix $\widetilde{\mathbf{H}}_{\mathrm{ind}}(\bm{\phi})$ and $\widetilde{\mathbf{H}}_{\mathrm{ind},\mathrm{c}}(\bm{\phi})$, which hinders the solution of the problem (\ref{eq:FP transformation}) for $\bm{\phi}$,
we need to perform some equivalent transformations on $\widetilde{\mathbf{H}}_{\mathrm{ind}}(\bm{\phi})\mathbf{x}$ and $\widetilde{\mathbf{H}}_{\mathrm{ind},\mathrm{c}}(\bm{\phi})\mathbf{x}$ to obtain more explicit objective function expressions for $\bm{\phi}$.
Firstly, we construct a simplified form for $\mathbf{H}_{\mathrm{t},r,i}(\bm{\phi})$ and $\mathbf{H}_{q,r,i}(\bm{\phi})$.
Specially, since the reflection coefficient matrix $\bm{\Phi}_r$ is diagonal, we have $\mathbf{b}^T(\theta_{\mathrm{t},r})\bm{\Phi}_r= \bm{\phi}_r^T\mathrm{diag}\{\mathbf{b}(\theta_{\mathrm{t},r})\}$. Moreover, we define following matrices for brevity
\begin{align}\label{eq:define B}
\mathbf{B}_{\mathrm{t},r}&\!\triangleq\! \mathrm{diag}\{\mathbf{b}(\theta_{\mathrm{t},r})\}\mathbf{G}_r, \ \mathbf{B}_{q,r}\! \triangleq \!\mathrm{diag}\{\mathbf{b}(\theta_{q,r})\}\mathbf{G}_r.
\end{align}

By substituting (\ref{eq:define B}) into formulas (\ref{eq:define Htri}) and (\ref{eq:define Hqri}) respectively, we can obtain the simplified forms for $\mathbf{H}_{\mathrm{t},r,i}(\bm{\phi})$ and $\mathbf{H}_{q,r,i}(\bm{\phi})$.
Then, recalling the definition of $\widetilde{\mathbf{H}}_{\mathrm{ind}}(\bm{\phi})$ in equation (\ref{eq:define Hdir Hind}) and utilizing the properties of Kronecker product,
an alternative equivalent form for each indirect path can be expressed as
\begin{subequations}\begin{align}
&\big[\mathbf{D}(f_{\text{d},r,1})\otimes\mathbf{J}^T_{\text{t},r,1}\otimes\mathbf{H}_{\text{t},r,1}(\bm{\phi}_r)\big]\mathbf{x} \\
    &\quad= \alpha_{\text{t},r,1}\text{vec}\big\{\mathbf{B}_{\text{t},r}^T\bm{\phi}_r\mathbf{a}^T(\theta_\text{t})\mathbf{X}(\mathbf{D}(f_{\text{d},r,1})\otimes\mathbf{J}_{\text{t},r,1})\big\}\\
    &\quad =\alpha_{\text{t},r,1}\big([(\mathbf{D}(f_{\text{d},r,1})\otimes\mathbf{J}^T_{\text{t},r,1})\mathbf{X}^T\mathbf{a}(\theta_\text{t})]\otimes\mathbf{B}_{\text{t},r}^T\big)\bm{\phi}_r,\\
&\big[\mathbf{D}(f_{\text{d},r,2})\otimes\mathbf{J}^T_{\text{t},r,2}\otimes\mathbf{H}_{\text{t},r,2}(\bm{\phi}_r)\big]\mathbf{x} \\
    &\quad =\alpha_{\text{t},r,2}\big([(\mathbf{D}(f_{\text{d},r,2})\otimes\mathbf{J}^T_{\text{t},r,2})\mathbf{X}^T\mathbf{B}_{\text{t},r}^T]\otimes\mathbf{a}(\theta_\text{t})\big)\bm{\phi}_r,\\
&\big[\mathbf{D}(f_{\text{d},r,3})\otimes\mathbf{J}^T_{\text{t},r,3}\otimes\mathbf{H}_{\text{t},r,3}(\bm{\phi}_r)\big]\mathbf{x} \\
    &\quad =\alpha_{\text{t},r,3}\big([(\mathbf{D}(f_{\text{d},r,3})\otimes\mathbf{J}^T_{\text{t},r,3})\mathbf{X}^T\mathbf{B}_{\text{t},r}^T]\otimes\mathbf{B}_{\text{t},r}^T\big)\bm{\varphi}_r,
\end{align}\end{subequations}
where for conciseness we define
 \begin{align}
 \bm{\varphi}_r\triangleq \mathrm{vec}(\bm{\phi}_r\bm{\phi}_r^T)=\bm{\phi}_r \otimes \bm{\phi}_r.
 \end{align}
Then, we can obtain the following equivalent formula:
  \begin{equation}\label{eq:transformation for yt}
 \begin{aligned}
 \widetilde{\mathbf{H}}_{\mathrm{ind}}(\bm{\phi})\mathbf{x}&=\sum_{r \in \mathcal{R}}(\mathbf{F}_{\mathrm{t},r} \bm{\phi}_r+\widetilde{\mathbf{F}}_{\mathrm{t},r}\bm{\varphi}_r)\\
 &= \mathbf{F}_\mathrm{t} \bm{\phi}+\widetilde{\mathbf{F}}_\mathrm{t}\bm{\varphi},
 \end{aligned}
 \end{equation}
 where we define
\begin{subequations}\begin{align}
\mathbf{F}_{\mathrm{t},r}&\triangleq\alpha_{\text{t},r,1}\big([(\mathbf{D}(f_{\text{d},r,1})\otimes\mathbf{J}^T_{\text{t},r,1})\mathbf{X}^T\mathbf{a}(\theta_\text{t})]\otimes\mathbf{B}_{\text{t},r}^T\big)\non\\
&\quad+\alpha_{\text{t},r,2}\big([(\mathbf{D}(f_{\text{d},r,2})\!\otimes\!\mathbf{J}^T_{\text{t},r,2})\mathbf{X}^T\mathbf{B}_{\text{t},r}^T]\otimes\mathbf{a}(\theta_\text{t})\big),\\
\widetilde{\mathbf{F}}_{\mathrm{t},r}&\triangleq\alpha_{\text{t},r,3}\big([(\mathbf{D}(f_{\text{d},r,3})\otimes\mathbf{J}^T_{\text{t},r,3})\mathbf{X}^T\mathbf{B}_{\text{t},r}^T]\otimes\mathbf{B}_{\text{t},r}^T\big),\\
\mathbf{F}_\mathrm{t}&\triangleq [\mathbf{F}_{\mathrm{t},1}, \cdots, \mathbf{F}_{\mathrm{t},R}],\quad
\widetilde{\mathbf{F}}_\mathrm{t}\triangleq [\widetilde{\mathbf{F}}_{\mathrm{t},1}, \cdots, \widetilde{\mathbf{F}}_{\mathrm{t},R}],\\
\bm{\varphi}&\triangleq [\bm{\varphi}_1^T,\bm{\varphi}_2^T,\ldots,\bm{\varphi}_R^T]^T
\end{align}\end{subequations}
to enable a more compact form of the reflection coefficients for all active RISs.
Following the similar procedure, we have
\begin{equation}\label{eq:transformation for yc}
\widetilde{\mathbf{H}}_{\mathrm{ind},\mathrm{c}}(\bm{\phi})\mathbf{x} = \mathbf{F}_\text{c}\bm{\phi}+\widetilde{\mathbf{F}}_\text{c}\bm{\varphi}.
\end{equation}
We omit the detailed definitions of $\mathbf{F}_\text{c}$ and $\widetilde{\mathbf{F}}_\text{c}$ here for brevity.

Substituting \eqref{eq:transformation for yt} and \eqref{eq:transformation for yc} into problem (\ref{eq:FP transformation}), we can obtain the
objective function $f_2(\bm{\phi})$ as
\begin{align}\label{eq:define f2}
f_2(\bm{\phi})&=\bm{\phi}^H\mathbf{C}\bm{\phi}+ \Re\{\bm{\phi}^H\mathbf{c}\}+\bm{\varphi}^H\widetilde{\mathbf{C}}\bm{\varphi}\\ \nonumber
&\qquad+ \Re\{\bm{\varphi}^H\widetilde{\mathbf{c}}\}+\Re\{\bm{\varphi}^H\overline{\mathbf{C}}\bm{\phi}\}+c_2,
\end{align}
where for brevity we define
\begin{subequations}
\begin{align}
\mathbf{C} &\triangleq \eta \mathbf{F}_\text{c}^H\mathbf{w}\mathbf{w}^H\mathbf{F}_\text{c}-\mathbf{F}_\mathrm{t}^H\mathbf{w} \mathbf{w}^H\mathbf{F}_\mathrm{t},\\
\widetilde{\mathbf{C}}&\triangleq    \eta \widetilde{\mathbf{F}}_\text{c}^H\mathbf{w}\mathbf{w}^H \widetilde{\mathbf{F}}_\text{c}-\widetilde{\mathbf{F}}_\mathrm{t}^H\mathbf{w} \mathbf{w}^H\widetilde{\mathbf{F}}_\mathrm{t},\\
\mathbf{c}  &\triangleq   2\eta \mathbf{F}_\text{c}^H\mathbf{w}\mathbf{w}^H\widetilde{\mathbf{H}}_{\mathrm{dir},\mathrm{c}}\mathbf{x}-2\mathbf{F}_\mathrm{t}^H\mathbf{w} \mathbf{w}^H\widetilde{\mathbf{H}}_{\mathrm{dir}}\mathbf{x},\\
\widetilde{\mathbf{c}}& \triangleq   2\eta \widetilde{\mathbf{F}}_\text{c}^H\mathbf{w}\mathbf{w}^H\widetilde{\mathbf{H}}_{\mathrm{dir},\mathrm{c}}\mathbf{x}-2\widetilde{\mathbf{F}}_{\mathrm{t}}^H\mathbf{w} \mathbf{w}^H\widetilde{\mathbf{H}}_{\mathrm{dir}}\mathbf{x},\\
\overline{\mathbf{C}}&\triangleq   2\eta \widetilde{\mathbf{F}}_\text{c}^H\mathbf{w}\mathbf{w}^H\mathbf{F}_\text{c}-2\widetilde{\mathbf{F}}_{\mathrm{t}}^H\mathbf{w} \mathbf{w}^H\mathbf{F}_\mathrm{t},\\
c_2& \triangleq  \eta |\mathbf{w}^H\widetilde{\mathbf{H}}_{\mathrm{dir},\mathrm{c}}\mathbf{x}|^2-|\mathbf{w}^H\widetilde{\mathbf{H}}_{\mathrm{dir}}\mathbf{x}|^2.
\end{align}
\end{subequations}
Then the optimization problem with respect to $\bm{\phi}$ can be rewritten as
\begin{subequations}\label{eq:problem for phi}
\begin{align}
\min_{\bm{\phi}}&\ \ f_2(\bm{\phi})\\
\text{s.t.}&\ \    \Re\{\widetilde{\mathbf{h}}_i^T(\bm{\phi})\mathbf{x}\}\geq \gamma_i,~\forall i,\\
&\ \ |\phi_n|\leq a_\mathrm{max},~\forall n.
\end{align}
\end{subequations}
It can be observed that the objective function contains quartic term $\bm{\varphi}^H\widetilde{\mathbf{C}}\bm{\varphi}$, cubic term $\Re\{\bm{\varphi}^H\overline{\mathbf{C}}\bm{\phi}\}$, and non convex quadratic terms $\bm{\phi}^H\mathbf{C}\bm{\phi}$ and $\Re\{\bm{\varphi}^H\widetilde{\mathbf{c}}\}$, making it an extremely complex non convex objective function.
In addition, the variable $\bm{\phi}$ is embedded in the function $\widetilde{\mathbf{h}}_i(\bm{\phi})$, which is not amenable for optimization.
Therefore, some relaxation of the objective function (\ref{eq:problem for phi}a) and mathematical transformations of $\widetilde{\mathbf{h}}_i(\bm{\phi})$ are very necessary.

For the sake of addressing the non-convex terms in the objective function, we try to construct a series of upper bound as the surrogate function that approximates the complicated non-convex term in each iteration.
Specially, the surrogate function of the non-convex term $-\bm{\phi}^H\mathbf{F}_\mathrm{t}^H\mathbf{w}\mathbf{w}^H\mathbf{F}_\mathrm{t}\bm{\phi}$ in $\bm{\phi}^H\mathbf{C}\bm{\phi}$ is constructed as
\begin{equation}\label{eq:negative term for phi}
-\bm{\phi}^H\mathbf{F}_\mathrm{t}^H\mathbf{w}\mathbf{w}^H\mathbf{F}_\mathrm{t}\bm{\phi}\leq-2\Re\{\bm{\phi}^H\mathbf{F}_\mathrm{t}^H\mathbf{w}\mathbf{w}^H\mathbf{F}_\mathrm{t}\bm{\phi}^{(t)}\}+c_4,
\end{equation}
where $\bm{\phi}^{(t)}$ is the solution to $\bm{\phi}$ obtained in the previous iteration and $c_4 \triangleq (\bm{\phi}^{(t)})^H \mathbf{F}_\mathrm{t}^H\mathbf{w}\mathbf{w}^H\mathbf{F}_\mathrm{t}\bm{\phi}^{(t)}$.
As for the non-convex term $\bm{\varphi}^H\widetilde{\mathbf{C}}\bm{\varphi}$, it is a quartic function about $\bm{\phi}$, so we firstly use the second-order Taylor expansion of $\bm{\varphi}$ to approximate it as
\begin{align}
\bm{\varphi}^H\widetilde{\mathbf{C}}\bm{\varphi}&\leq \lambda_1 \bm{\varphi}^H\bm{\varphi}+2\Re \{\bm{\varphi}^H(\widetilde{\mathbf{C}}-\lambda_1\mathbf{I}_{RN_\mathrm{r}^2})\bm{\varphi}^{(t)}\}\\ \nonumber
&\quad~ +(\bm{\varphi}^{(t)})^H(\lambda_1\mathbf{I}_{RN_\mathrm{r}^2}-\widetilde{\mathbf{C}})\bm{\varphi}^{(t)},
\end{align}
where $\lambda_1$ is an upper bound of the eigenvalues of matrix $\widetilde{\mathbf{C}}$.
The value of the auxiliary variable $\bm{\varphi}^{(t)}$ at the
$t$-th iteration is obtained based on $\bm{\phi}^{(t)}$.
To avoid the complexity of high-dimensional matrix eigenvalue decomposition, we use $\lambda_1=\mathrm{Tr}\{\widetilde{\mathbf{C}}\}$ as an alternative.
Besides, thanks to the amplitude constraint of active RIS (\ref{eq:problem for phi}c), $\lambda_1 \bm{\varphi}^H\bm{\varphi}$ is upper-bounded by
\begin{equation}
\lambda_1 \bm{\varphi}^H\bm{\varphi}\leq \lambda_1 RN_\mathrm{r}^2 a^4_\mathrm{max}.
\end{equation}
Then the quartic function $\bm{\varphi}^H\widetilde{\mathbf{C}}\bm{\varphi}$ of $\bm{\phi}$ is reduced to a quadratic term.
Considering that $\Re\{\bm{\varphi}^H\widetilde{\mathbf{c}}\}$ is also a quadratic term about $\bm{\phi}$, the overall surrogate function for $\bm{\varphi}^H\widetilde{\mathbf{C}}\bm{\varphi}+\Re\{\bm{\varphi}^H\widetilde{\mathbf{c}}\}$ can be represented as
\begin{equation}
\bm{\varphi}^H\widetilde{\mathbf{C}}\bm{\varphi}+\Re\{\bm{\varphi}^H\widetilde{\mathbf{c}}\}\leq \Re\{\bm{\varphi}^H\bm{\ell}\}+c_5,
\end{equation}
where we define $\bm{\ell}\triangleq 2\widetilde{\mathbf{C}}\bm{\varphi}^{(t)}+\widetilde{\mathbf{c}}$, and the constant $c_5$ is independent of $\bm{\phi}$.
Utilizing the properties of the Kronecker product, another equivalent form of $\Re\{\bm{\varphi}^H\bm{\ell}\}$ is
\begin{equation}
\begin{aligned}
\Re\{\bm{\varphi}^H\bm{\ell}\}&=\sum_{r\in\mathcal{R}}\Re\{\bm{\varphi}_r^H\bm{\ell}_r\}=\sum_{r\in\mathcal{R}}\Re\{\bm{\phi}_r^H\mathbf{L}_r\bm{\phi}_r^\ast\}\\
&=\Re\{\bm{\phi}^H\mathbf{L}\bm{\phi}^\ast\},
\end{aligned}
\end{equation}
where $\bm{\ell}_r\in\mathbb{C}^{N_\text{r}\times 1}$ is the $r$-th sub-vector of $\bm{\ell}$ and $\mathbf{L}_r$ is a reshaped version of $\bm{\ell}_r$ with $\bm{\ell}_r= \mathrm{vec}\{\mathbf{L}_r\}$.
Besides, the matrix $\mathbf{L}$ is a block diagonal matrix defined by $\mathbf{L}\triangleq \mathrm{blkdiag}\{\mathbf{L}_1,\mathbf{L}_2,\cdots,\mathbf{L}_R\} $.

Due to the non-convexity of the real-valued function $\Re\{\bm{\phi}^H\mathbf{L}\bm{\phi}^\ast\}$, we propose to rewrite it as real-valued form and seek for a more tractable upper bound via the second-order Taylor expansion.
In particular, by defining
\begin{subequations}
\begin{align}
\overline{\bm{\phi}}&\triangleq [\Re\{\bm{\phi}^T\}~~ \Im\{\bm{\phi}^T\}]^T,\\
\overline{\mathbf{L}} &\triangleq\left[\begin{array}{ll}
\Re\left\{\mathbf{L}\right\} & \Im\left\{\mathbf{L}\right\} \\
\Im\left\{\mathbf{L}\right\} & -\Re\left\{\mathbf{L}\right\}
\end{array}\right],
\end{align}
\end{subequations}
the convex surrogate function is constructed as
\begin{subequations}\label{eq:real valued version of quartic function}
\begin{align}
\Re\{\bm{\phi}^H\mathbf{L}\bm{\phi}^\ast\}&= \overline{\bm{\phi}}^T\overline{\mathbf{L}}\overline{\bm{\phi}}\\
&\leq  (\overline{\bm{\phi}}^{(t)})^T\overline{\mathbf{L}}\overline{\bm{\phi}}^{(t)}+(\overline{\bm{\phi}}^{(t)})^T(\overline{\mathbf{L}}+\overline{\mathbf{L}}^T)(\overline{\bm{\phi}}-\overline{\bm{\phi}}^{(t)}) \nonumber\\
&\quad~+\frac{\lambda_2}{2}(\overline{\bm{\phi}}-\overline{\bm{\phi}}^{(t)})^T(\overline{\bm{\phi}}-\overline{\bm{\phi}}^{(t)})\\
&=\frac{\lambda_2}{2} \bm{\phi}^H \bm{\phi}+\Re\{\bm{\phi}^H\mathbf{U}\overline{\bm{\ell}}\}+c_6,
\end{align}
\end{subequations}
in which $\lambda_2$ is the maximum eigenvalue of the Hessian matrix $(\overline{\mathbf{L}}+\overline{\mathbf{L}}^T)$, $\overline{\bm{\ell}}\triangleq (\overline{\mathbf{L}}+\overline{\mathbf{L}}^T-\lambda_2\mathbf{I}_{2RN_\mathrm{r}})\overline{\bm{\phi}}^{(t)}$ and $\mathbf{U}\triangleq [\mathbf{I}_{RN_\mathrm{r}}\  \jmath\mathbf{I}_{RN_\mathrm{r}}]$.
The constant scalar $c_6$ is unrelated to $\bm{\phi}$ with the definition $c_6 \triangleq -(\overline{\bm{\phi}}^{(t)})^T \overline{\mathbf{L}}^T \overline{\bm{\phi}}^{(t)}+\lambda_2/2(\overline{\bm{\phi}}^{(t)})^T\overline{\bm{\phi}}^{(t)}$.

The non-convex term $\Re\{\bm{\varphi}^H\overline{\mathbf{C}}\bm{\phi}\}$ in the objective function $f_2(\bm{\phi})$ is a cubic term with respect to $\bm{\phi}$, which poses a significant challenge in solving the problem.
We propose to utilize the second-order Taylor expansion at the iteration point $\bm{\phi}^{(t)}$ to obtain a more tractable upper bound as the surrogate function.
Specifically, the surrogate function for $\Re\{\bm{\varphi}^H\overline{\mathbf{C}}\bm{\phi}\}$ can be expressed as
\begin{equation}\label{eq:cubic term}
\begin{aligned}
\Re\{\bm{\varphi}^H\overline{\mathbf{C}}\bm{\phi}\}
&\leq \Re\{(\bm{\varphi}^{(t)})^H\overline{\mathbf{C}}\bm{\phi}^{(t)}\}+\frac{1}{2}\Re\big\{(\bm{\phi}-\bm{\phi}^{(t)})^H[\overline{\mathbf{C}}^H\bm{\varphi}^{(t)}\\
&\!+\!(\bm{\Xi}^{(t)})^H\overline{\mathbf{C}}\bm{\phi}^{(t)}]\big\}\!+\!\frac{\lambda_3}{2}(\bm{\phi}\!-\!\bm{\phi}^{(t)})^H(\bm{\phi}\!-\!\bm{\phi}^{(t)})\\
&=\frac{\lambda_3}{2}\bm{\phi}^H\bm{\phi}+\Re\{\bm{\phi}^H\overline{\mathbf{c}}\}+c_7,
\end{aligned}
\end{equation}
\nid where the matrix $\bm{\Xi}\!\triangleq\! \mathrm{blkdiag}\{\bm{\Xi}_1, \bm{\Xi}_2, \cdots, \bm{\Xi}_R\}$ is an auxiliary matrix introduced during the differentiation process with $\bm{\Xi}_r\triangleq \mathbf{I}_{N_\mathrm{r}}\!\otimes\bm{\phi}_r \!+\!\bm{\phi}_r \!\otimes\mathbf{I}_{N_\mathrm{r}}$.
The value of $\bm{\Xi}^{(t)}$ can be obtained according to $\bm{\phi}_r^{(t)}, ~\forall r$.
The scalar $\lambda_3$ represents the maximum eigenvalue of the Hessian matrix $\frac{1}{2}(\overline{\mathbf{C}}^H\bm{\Xi}^{(t)}+(\bm{\Xi}^{(t)})^H\overline{\mathbf{C}})$.
For conciseness, we define
$\overline{\mathbf{c}}\!\triangleq\!\frac{1}{2}(\overline{\mathbf{C}}^H\bm{\varphi}^{(t)}\!+\!(\bm{\Xi}^{(t)})^H\overline{\mathbf{C}}\bm{\phi}^{(t)})$, and the scalar $c_7$ is independent of $\bm{\phi}$.
Substituting (\ref{eq:negative term for phi}), (\ref{eq:real valued version of quartic function}) and
 (\ref{eq:cubic term}) into (\ref{eq:define f2}), we obtain the overall upper bound of $f_2(\bm{\phi})$ as
\begin{equation}
f_2(\bm{\phi})\leq \bm{\phi}^H\mathbf{M}\bm{\phi}+\Re\{\bm{\phi}^H\mathbf{m}\}+c_8,
\end{equation}
where we define
\begin{subequations}
\begin{align}
\mathbf{M}&\triangleq \eta\mathbf{F}_\text{c}^H\mathbf{w}\mathbf{w}^H\mathbf{F}_\text{c} + \frac{\lambda_2+\lambda_3}{2}\mathbf{I}_{RN_\text{r}},\\
\mathbf{m}&\triangleq -2\mathbf{F}_\mathrm{t}^H\mathbf{w}\mathbf{w}^H\mathbf{F}_\mathrm{t}\bm{\phi}^{(t)}+\mathbf{U}\overline{\bm{\ell}}+\mathbf{c}+\overline{\mathbf{c}},\\
c_8&\triangleq c_4+c_5+c_6+c_7.
\end{align}
\end{subequations}

After solving the non-convex objective function (\ref{eq:problem for phi}a), we consider the mathematical transformation of $\widetilde{\mathbf{h}}_i(\bm{\phi})$ to obtain an intuitive form of $\bm{\phi}$.
Combining the definition of $\widetilde{\mathbf{h}}_i(\bm{\phi})$ in formula (\ref{eq:define hitilde phi}) with the properties of Kronecker product, we can obtain
\begin{equation}\label{eq: relationship for hitilde x}
(\mathbf{e}_{j,ML}^T\otimes\mathbf{h}^T_{k}(\bm{\phi}))\mathbf{x}= \mathbf{h}^T_k(\bm{\phi})\mathbf{X}\mathbf{e}_{j,ML}.
\end{equation}
Substituting (\ref{eq:define hk xi}) and (\ref{eq: relationship for hitilde x}) into
$\Re\{\widetilde{\mathbf{h}}^T_i(\bm{\phi})\mathbf{x}\}$, we can acquire an equivalent form for $\Re\{\widetilde{\mathbf{h}}^T_i(\bm{\phi})\mathbf{x}\}$ as
\begin{equation}
\Re\{\widetilde{\mathbf{h}}^T_i(\bm{\phi})\mathbf{x}\}=\Re\{d_i+\bm{\phi}^T\mathbf{g}_i\},
\end{equation}
where we define
\begin{subequations}
\begin{align}
d_{(2k-1)ML+j} & \triangleq \xi_{m,k,1}\mathbf{h}_{\text{d},k}^T\mathbf{X}\mathbf{e}_{j,ML},\\
d_{(2k-2)ML+j} & \triangleq \xi_{m,k,2}\mathbf{h}_{\text{d},k}^T\mathbf{X}\mathbf{e}_{j,ML},\\
\mathbf{g}_{(2k-2)ML+j,r}&\triangleq \xi_{m,k,1}\mathrm{diag}\{\mathbf{h}_{\mathrm{r},k}\}\mathbf{G}_r\mathbf{X}\mathbf{e}_{j,ML},\\
\mathbf{g}_{(2k-2)ML+j,r}&\triangleq \xi_{m,k,2}\mathrm{diag}\{\mathbf{h}_{\mathrm{r},k}\}\mathbf{G}_r\mathbf{X}\mathbf{e}_{j,ML},\\
\mathbf{g}_{i}&\triangleq [\mathbf{g}^T_{i,1},\mathbf{g}^T_{i,2}, \cdots, \mathbf{g}^T_{i,R}]^T.
\end{align}
\end{subequations}
Then the constraints on communication QoS can be reformulated as
\begin{equation}
\Re\{d_i+\bm{\phi}^T\mathbf{g}_i\}\geq \gamma_i, ~\forall i,
\end{equation}
which is a specific expression for $\bm{\phi}$.

After the relaxation and equivalent transformation mentioned above,
and with fixed receive filter $\mathbf{w}$ and transmit waveform $\mathbf{x}$, the optimization problem with respect to $\bm{\phi}$ is
\begin{subequations}\label{eq:update phi}
\begin{align}
\min_{\bm{\phi}} & \ \ \bm{\phi}^H\mathbf{M}\bm{\phi}+\Re\{\bm{\phi}^H\mathbf{m}\}\\
\text{s.t.}& \ \ \Re\{d_i+\bm{\phi}^T\mathbf{g}_i\}\geq \gamma_i, ~\forall i,\\
&\ \ |\phi_n|\leq a_\mathrm{max}, ~\forall n.
\end{align}
\end{subequations}
Obviously, it is a convex problem which can be easily solved by various convex toolboxs.

\begin{algorithm}[!t]
\begin{small}
\caption{Joint Receive Filter, Transmit Waveform and Active RIS Reflection Design Algorithm}
\label{alg1}
    \begin{algorithmic}[1]
    \REQUIRE $\mathbf{H}_{\mathrm{t},0}$, $\mathbf{H}_{q,0}, \forall q$, $\mathbf{B}_{\mathrm{t},r}, \forall r$, $\mathbf{B}_{q,r},  \forall q, \forall r$, $\mathbf{h}_{\mathrm{d},k}, \forall k$, $\mathbf{h}_{\mathrm{r},k}, \forall k$, $\mathbf{G}$, $\Gamma$, $\sigma^2_\mathrm{z}$ $\sigma^2_{k}$, $a_\mathrm{max}$ and $P_\mathrm{BS}$.
    \ENSURE $\mathbf{w}^\star$, $\mathbf{x}^\star$ and $\bm{\phi}^\star$.
        \STATE {Initialize $\bm{\phi}$, $\mathbf{x}$, $\mathbf{w}$ and $\eta$ by (\ref{eq:initialization phi}), (\ref{eq:initialization x}), (\ref{eq:optimal w}) and (\ref{eq: update auxiliary variable eta}), respectively, and $\bm{\psi}:=\mathbf{x}$, $\bm{\lambda}:=\mathbf{0}$.}
        \WHILE {no convergence }
            \STATE{Update $\mathbf{x}$ by solving (\ref{eq:ADMM update x}).}
            \STATE{Update $\bm{\psi}$ using (\ref{eq: closed form for y}).}
            \STATE{Update $\bm{\lambda}$ using (\ref{eq:ADMM update lambda}).}
            \STATE{Update $\bm{\phi}$ by solving (\ref{eq:update phi}).}
            \STATE{Update $\mathbf{w}$ using (\ref{eq:optimal w}).}
            \STATE{Update $\eta$ using (\ref{eq: update auxiliary variable eta}).}
        \ENDWHILE
        \STATE{Return $\mathbf{w}^\star =\mathbf{w}$, $\mathbf{x}^\star = \mathbf{x}$ and $\bm{\phi}^\star =\bm{\phi}$.}
    \end{algorithmic}
    \end{small}
\end{algorithm}

\subsection{Summary, Initialization, and Complexity Analysis}
Based on the above derivation, the proposed joint design algorithm for active RIS-aid ISAC system is straightforward. With appropriate initialization, conditionally optimal transmit waveform $\mathbf{x}$, receive filter $\mathbf{w}$, and RIS coefficient $\bm{\phi}$ are iteratively obtained until the convergence is achieved. The procedure of this proposed joint design algorithm is summarized in Algorithm 1.

To accelerate the convergence of the proposed algorithm, it is crucial to properly initialize $\bm{\phi}$, $\mathbf{w}$, and $\mathbf{x}$. The initialization process is outlined as follows.
Since the active RIS is employed to enhance the wireless propagation environment, the channel gains are utilized as the performance metric to determine the initial value of $\bm{\phi}$.
Specifically, ignoring the magnitude of the ARIS amplification coefficient and focusing solely on optimizing its phase, denoted as $\bm{\beta}\triangleq [\beta_1,\cdots,\beta_{N_\mathrm{r}},\cdots,\beta_{RN_\mathrm{r}}]^T$, the initialization problem is formulated as
\begin{subequations}\label{eq:initialization phi}
\begin{align}
\max_{\bm{\beta}}&\ \  \big\|\mathbf{a}^T(\theta_\mathrm{t})+\sum_{r \in \mathcal{R}}\bm{\beta}_r^T\mathbf{B}_{\mathrm{t},r}\big\|^2\\ \nonumber
&\quad - \sum_{q \in \mathcal{Q}}\big\|\mathbf{a}^T(\theta_{q})+\sum_{r \in \mathcal{R}}\bm{\beta}_r^T\mathbf{B}_{q,r}\big\|^2\\
\text{s.t.}&\ \ |\beta_j|=1,~\forall j,
\end{align}
\end{subequations}
which can be efficiently solved by the Riemannian conjugate gradient (RCG) algorithm \cite{Liu TWC 2021}. The initial value of the reflection coefficients is then computed as $\bm{\phi}=a_\mathrm{max}\bm{\beta}$.
Following the initialization of $\bm{\phi}$, the transmit waveform $\mathbf{x}$ is initialized by maximizing the QoS for the worst-case communication user. This leads to the following optimization problem for initializing $\mathbf{x}$
\begin{subequations}\label{eq:initialization x}
\begin{align}
\max_{\mathbf{x}} &\ \ \delta\\
\text{s.t.}&\ \ \Re\{\widetilde{\mathbf{h}}_i^T(\bm{\phi})\mathbf{x}\}\geq \delta,~ \forall i,\\
&\ \ |x_j|\leq \sqrt{P_\mathrm{BS}/(NML)}, ~\forall j.
\end{align}
\end{subequations}
Finally, given the initialized values of
$\bm{\phi}$ and $\mathbf{x}$, the initialization of receive filter $\mathbf{w}$ and the auxiliary variable $\eta$ can be obtained using (\ref{eq:optimal w}) and (\ref{eq: update auxiliary variable eta}), respectively.

Next, we provide a brief analysis of the computational complexity of the proposed algorithm. The complexity of solving convex optimization problems using the interior-point method is assumed to depend on the dimensionality of the optimization variables as well as the number of linear and quadratic constraints.
During the initialization stage, the computational complexities of obtaining the initial values of $\bm{\phi}$ and $\mathbf{x}$ are $\mathcal{O}\{(RN_\mathrm{r})^{1.5}\}$ and $\mathcal{O}\{\sqrt{NML+2KML}(NML)^3\}$, respectively.
In the iterative process, updating the waveform $\mathbf{x}$ by solving problem (\ref{eq:ADMM update x}) incurs a computational complexity of
 $\mathcal{O}\{\sqrt{NML+2KML}(NML)^3\}$.
Solving problem (\ref{eq:problem for phi}) to update the reflections $\bm{\phi}$ has the computational complexity of $\mathcal{O}\{\sqrt{RN_\mathrm{r}+2KML}(RN_\mathrm{r})^3\}$.
The updates of the auxiliary variable $\bm{\psi}$  and dual variable $\bm{\lambda}$ through (\ref{eq:ADMM update y}) and (\ref{eq:ADMM update lambda}), respectively, each have a computational complexity $\mathcal{O}\{NML\}$. Lastly, updating the receive filter using the closed-form expression in (\ref{eq:optimal w}) requires $\mathcal{O}\{NMP\}$ operations.
Thus, the overall computational complexity is of order $\mathcal{O}\{\sqrt{NML+2KML}(NML)^3+\sqrt{RN_\mathrm{r}+2KML}(RN_\mathrm{r})^3\}$.

\section{Simulation Results}\label{sec:simulation results}

\begin{figure}[!t]
\centering
\includegraphics[width = 3.0 in]{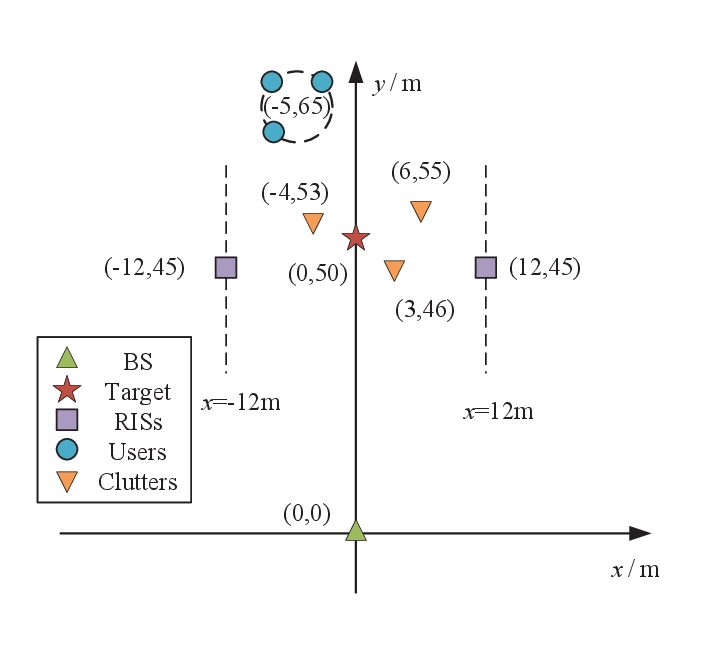}\vspace{-0.5 cm}
\caption{An illustration of the positions of BS, RISs, users, target and clutters.}\label{fig:setting}
\end{figure}

In this section, we present extensive simulation results to demonstrate the effectiveness of the proposed joint design algorithm for transmit waveform, active RIS reflection, and receive filter optimization. Unless stated otherwise, the following conditions apply to all simulation experiments. As illustrated in Fig. \ref{fig:setting}, the dual-function BS is located at the origin of the coordinate system and is equipped with $N = 8$ transmit/receive antennas. The system serves $K = 3$ communication users, with positions within a 4-meter radius centered at $(-5\,\text{m}, 65\,\text{m})$. The target is positioned at $(0\,\text{m}, 50\,\text{m})$. Additionally, three static clutters are located at $(6\,\text{m}, 55\,\text{m})$, $(-4\,\text{m}, 53\,\text{m})$, and $(3\,\text{m}, 46\,\text{m})$, each within a 4-meter radius.
The ISAC system is assisted by $R = 2$ active RISs, located at $(-12\,\text{m}, 45\,\text{m})$ and $(12\,\text{m}, 45\,\text{m})$, each equipped with $N_\mathrm{r} = 25$ elements. The typical path-loss model $PL(d) = C_0(d_0/d)^\iota$ is adopted for the ISAC system, where the path-loss exponents for the target's direct path
fading, $\alpha_{\mathrm{t},0}$ and the clutters' direct path fading, $\alpha_{q,0}$ are set to 2.7, and the indirect path fading coefficients for the target $\alpha_{\mathrm{t},r,i}$ and the clutters $\alpha_{q,r,i}$ are set to 2.3.
The Rayleigh channel model is used for both the communication channel between the BS and the users, and the channel between the RIS and the users. The path-loss exponents for the channels $\mathbf{h}_{\mathrm{d},k}$, $\mathbf{h}_{r,k}$, and $\mathbf{G}_r$ are set to 3.0, 2.8, and 2.0, respectively. The communication QoS requirement is $\Gamma_k = 10$dB, $\forall k$. Each coherent pulse interval contains $M = 8$ pulses, with a PRF of $1000$ Hz, and each pulse contains $L = 8$ sample points. The carrier frequency of the transmit waveform is set to $f_0 = 2.4$ GHz. Additionally, the noise power is set to $\sigma_\mathrm{r}^2 = \sigma^2_k = -80$dB, $\forall k$.

\begin{figure}[!t]
\centering
\includegraphics[width = 3.5 in]{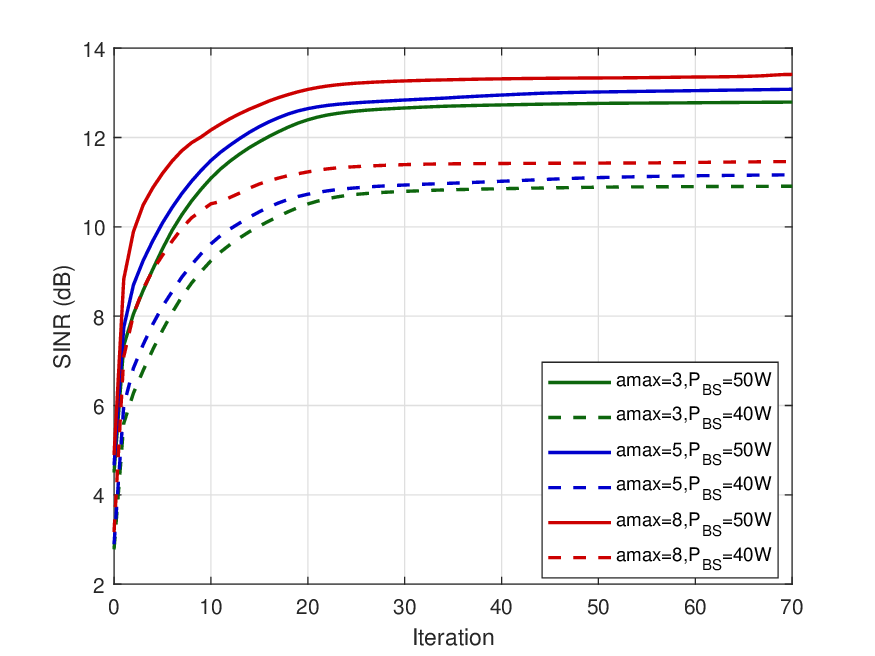}\vspace{0.0 cm}
\caption{Convergence illustration.}\label{fig:iterations}
\centering
\includegraphics[width = 3.5 in]{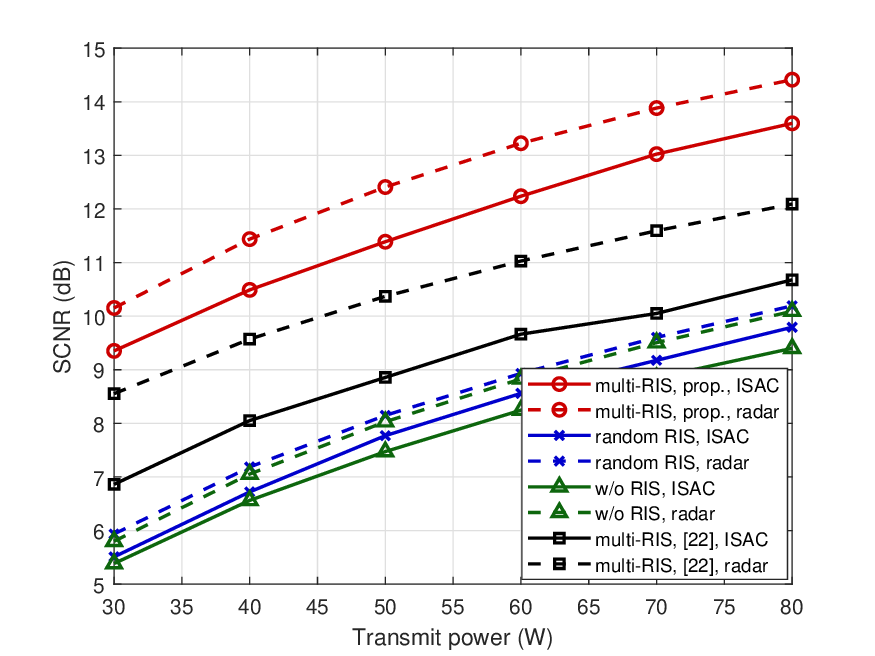}\vspace{0.0 cm}
\caption{Radar output SCNR versus transmit power.}\label{fig:SINR vs P}\vspace{-0.3 cm}
\end{figure}

The convergence performance of the proposed algorithm is illustrated in Fig. \ref{fig:iterations}. To evaluate the convergence behavior under different settings, the amplification coefficient $a_\mathrm{max}$ for the ARIS is chosen as 3, 5, and 8, while the transmission power for the BS is set to $50$W and $40$W, respectively. It is evident that the proposed algorithm demonstrates superior convergence performance, achieving convergence within a limited number of iterations across the considered scenarios.

Fig. \ref{fig:SINR vs P} illustrates the target SCNR as a function of the BS transmit power, where the amplification coefficient is fixed as $a_\mathrm{max} = 5$. To highlight the advantages of the proposed multi-perspective observation ISAC scheme (denoted as ``\textbf{multi-RIS, prop., ISAC}''), which leverages the spatial-delay-Doppler diversity of multiple controllable observation paths, we include three benchmark schemes for comparison: \textit{i}) the ISAC system employing STAP for clutter suppression, but with random phase-shift RIS (denoted as ``\textbf{random RIS, ISAC}''),  which utilizes multiple uncontrollable observation paths for sensing, \textit{ii}) the ISAC system employing STAP for clutter suppression, but without deploying RIS (denoted as ``\textbf{w/o RIS, ISAC}''), which utilizes single LoS observation path for sensing; \textit{iii}) and multi-RIS aided ISAC system \cite{Wei Tong ICOM 2023} (denoted as ``\textbf{multi-RIS, [22], ISAC}''), which utilizes multiple  observation paths but relies solely on spatial beamforming for clutter suppression. Additionally, to analyze the upper bound of detection performance, we evaluate the radar-only functionality of these schemes by excluding the QoS requirements for communication users. These radar-only scenarios are represented with dashed lines and labeled as ``\textbf{multi-RIS, prop., radar}'', ``\textbf{random RIS, radar}'', ``\textbf{w/o RIS, radar}'', and ``\textbf{multi-RIS, [22], radar}'', respectively.
The results presented in Fig. \ref{fig:SINR vs P} demonstrate that increasing the transmit power enhances radar output SCNR across all evaluated schemes. Notably, the proposed multi-perspective observation approach consistently outperforms its counterparts, achieving the highest radar SCNR performance in both ISAC and radar-only scenarios. This highlights the superior capability of the proposed approach in leveraging target characteristics across spatial, delay, and Doppler domains, thereby significantly improving target detection performance.

\begin{figure}[!t]
\centering
\includegraphics[width = 3.5 in]{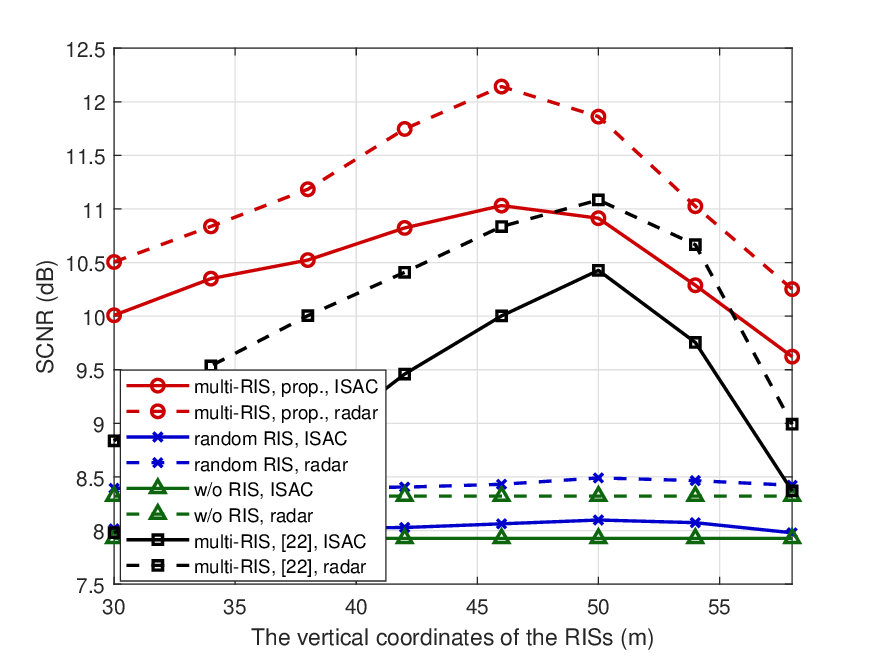}\vspace{0.0 cm}
\caption{Radar output SCNR versus the position of RISs.}\label{fig:SINR vs ARIS position}\vspace{-0.3 cm}
\end{figure}

Fig. \ref{fig:SINR vs ARIS position} illustrates the relationship between radar output SCNR and the deployment position of RISs. In this study, the horizontal coordinates of two RISs are fixed at $-12$m and $12$m, while their vertical coordinates vary within the range of $30$m to $58$m, i.e., along the vertical dashed line shown in Fig. \ref{fig:setting}.
From Fig. \ref{fig:SINR vs ARIS position}, it is evident that both the proposed multi-perspective observation approach and the spatial beamforming-based design \cite{Wei Tong ICOM 2023} achieve significantly improved, yet varying, SCNR performance. This variance arises because both schemes leverage multi-perspective observation by appropriately controlling the RISs, with the RIS positions having a substantial impact on sensing performance.
More importantly, the proposed multi-perspective observation approach demonstrates superior SCNR performance and reduced sensitivity to RIS deployment positions compared to the design in \cite{Wei Tong ICOM 2023}. This advantage and robustness stem from the scheme's ability to fully leverage spatial, delay, and Doppler characteristics of each observation path, in contrast to \cite{Wei Tong ICOM 2023}, which relies solely on spatial diversity through beamforming. These findings further underscore the adaptability and effectiveness of the proposed multi-perspective observation framework.

\begin{figure}[!t]
\centering
\includegraphics[width = 3.5 in]{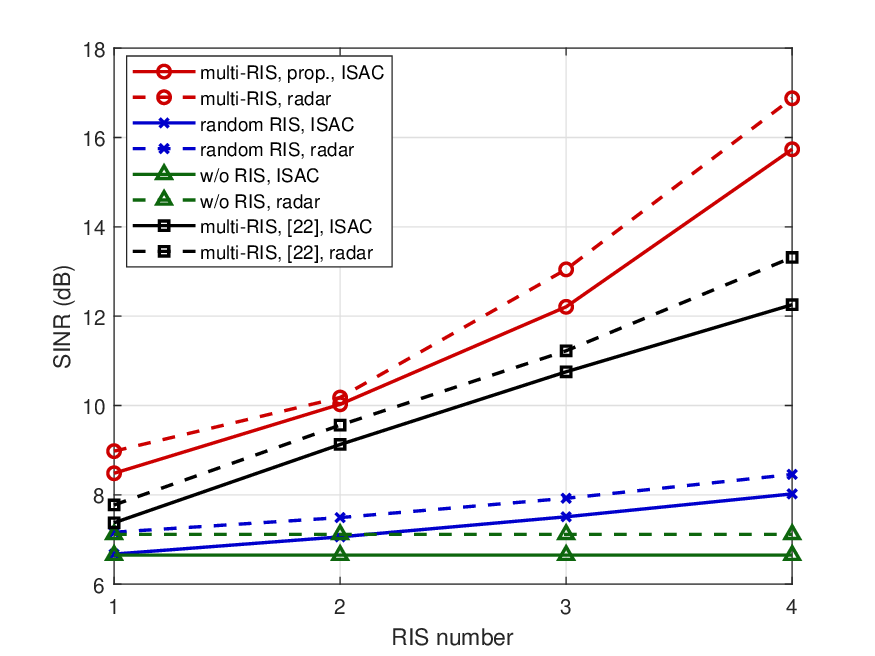}\vspace{0.0 cm}
\caption{Radar output SCNR versus the number of RISs.}\label{fig:SINR vs ARIS}\vspace{-0.3 cm}
\end{figure}

Furthermore, we evaluate the target SCNR as a function of the number of RISs, as shown in Fig. \ref{fig:SINR vs ARIS}. The results demonstrate that increasing the number of RISs consistently improves SCNR performance across all RIS-assisted schemes. Notably, the proposed multi-perspective observation approach achieves the most significant improvement. This improvement can be attributed to the more efficient utilization of multiple observation paths enabled by the additional RISs, which play a key role in mitigating clutter. These findings further corroborate the effectiveness of employing multiple RISs to enhance target detection performance in clutter-rich environments.

\begin{figure}[!t]
\centering
\includegraphics[width=3.5in]{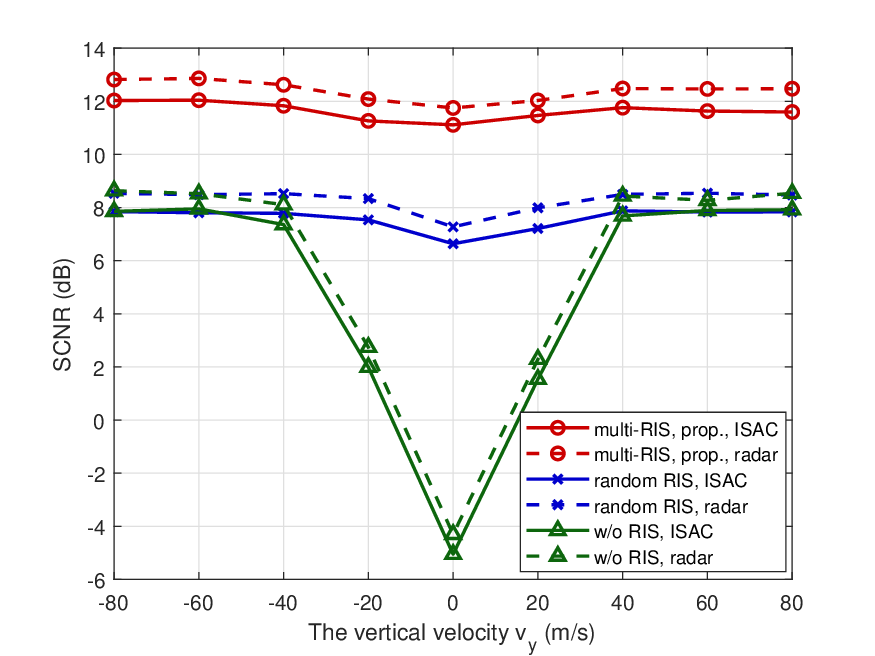}
\vspace{-0.3 cm}      \center{(a) SCNR versus the magnitude of the target's velocity.}
    \includegraphics[width=3.5in]{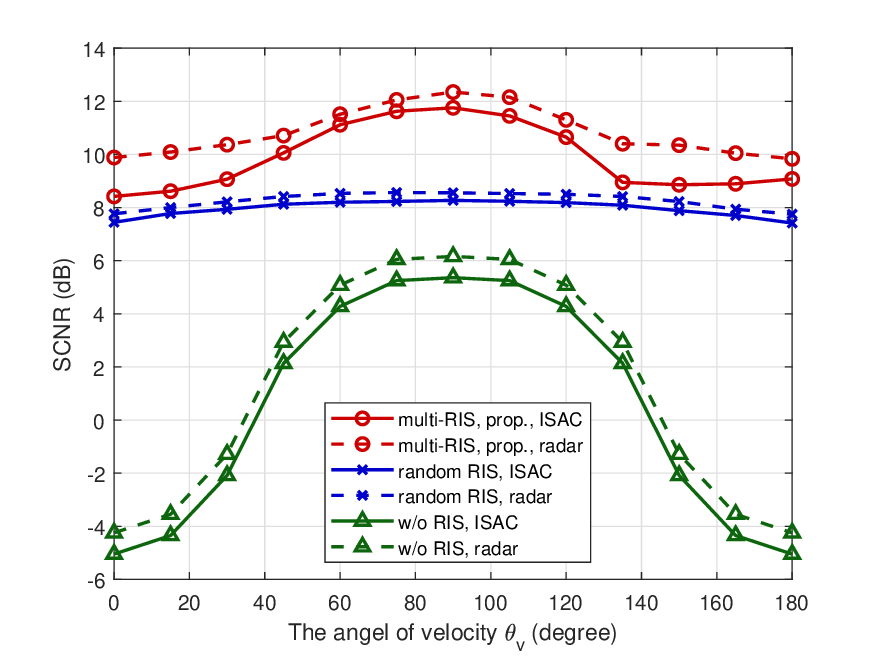}
    \vspace{-0.3 cm}   \center{(b) SCNR versus the direction of target's velocity.}
\caption{Radar output SCNR versus the target's velocity}
\label{fig:SINR vs velocity}
\vspace{-0.3cm}
\end{figure}

Previous simulation results demonstrate that the improvement provided by the proposed multi-perspective observation scheme capitalizes on the diversity of multiple observation paths, each with distinct radar cross-section (RCS), spatial, delay, and Doppler characteristics for both the target and clutter. To further illustrate this fact, Fig. \ref{fig:SINR vs velocity} presents an evaluation of target SCNR across varying target velocities.
Specifically, Fig. \ref{fig:SINR vs velocity}-(a) shows the relationship between radar output SCNR and the target velocity along the vertical direction, denoted as $v_y$. The results reveal that in schemes relying solely on the LoS path (i.e., without RIS), radar output SCNR significantly degrades at low target velocities due to interference from static clutter. In such cases, low relative velocities make it challenging to distinguish the target from clutter.
In contrast, the proposed multi-perspective observation strategy actively manipulates the propagation environment to generate multiple high-quality observation paths. By exploiting the diversity of these paths, the scheme effectively mitigates the challenges associated with single-path reliance, maintaining SCNR at a high level even at low velocities.
Moreover, Fig. \ref{fig:SINR vs velocity}-(b) illustrates the radar output SCNR versus the target's velocity direction. In this simulation, the target's velocity is fixed at 30 m/s, while its direction varies from $0^\circ$ to $180^\circ$. The results show that all three schemes achieve highest SCNR at $90^\circ$, corresponding to the LoS direction. For target movements deviating from the LoS direction, schemes relying solely on the LoS observation path encounter reduced relative target velocity along the vertical LoS direction. This reduction hinders the differentiation between the moving target and static clutter, leading to lower SCNR. The deployment of multiple RISs addresses this limitation by introducing additional observation paths, effectively mitigating the SCNR degradation by exploiting the diversity of multiple observation paths. For example, when the target moves horizontally, the LoS observation exhibits zero relative velocity, whereas the NLOS path via an RIS retains a non-zero velocity that closely approximates the target’s true motion, provided the RIS is positioned horizontally relative to the target. Therefore, the results presented in Fig. \ref{fig:SINR vs velocity} confirm that the proposed multi-perspective observation approach consistently maintains high SCNR across a wide range of velocities and directions, demonstrating its effectiveness in enhancing target detection performance in complex and clutter-rich environments.

\begin{figure}[!t]
\centering
\includegraphics[width = 3.5 in]{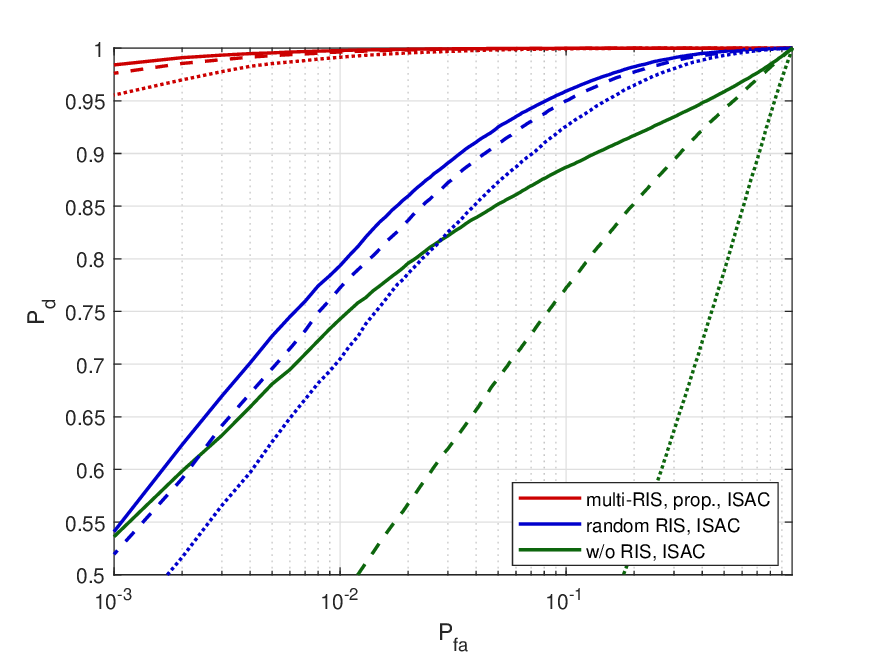}\vspace{0.0 cm}
\caption{Radar detection probability
versus false alarm probability. Solid lines: 60 m/s; dashed lines: 30 m/s; dotted lines 0 m/s.}\label{fig:ROC}
\vspace{0.0 cm}
\centering
\includegraphics[width = 3.5 in]{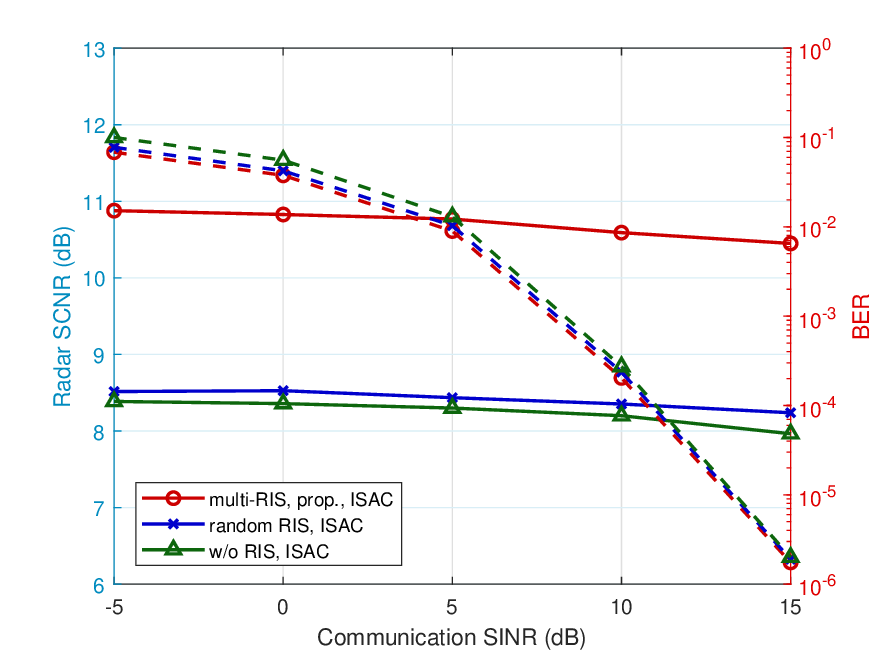}\vspace{0.0 cm}
\caption{Radar SCNR and communication BER versus communication SINR requirement.}\label{fig:SINR vs QoS}\vspace{-0.3 cm}
\end{figure}

To explicitly evaluate target detection performance, Fig. \ref{fig:ROC} presents the receiver operating characteristic (ROC) curve, plotting the detection probability ($P_{\mathrm{d}}$) against the false alarm probability ($P_{\mathrm{fa}}$). Solid lines correspond to a target velocity of 60 m/s, dashed lines to 30 m/s, and dotted lines to 0 m/s. The results demonstrate that the proposed multi-perspective observation approach consistently outperforms the two benchmark schemes in detection probability across varying target velocities, highlighting its superior target detection performance.

Finally, to evaluate the performance trade-off between radar sensing and communications in ISAC systems, Fig. \ref{fig:SINR vs QoS} illustrates the radar SCNR and communication bit error rate (BER) as functions of the communication SINR requirement $\gamma_i$. The solid lines represent the SCNR performance for  sensing, while the dashed lines correspond to the BER performance for communication functions. As expected, the BER performance of all schemes improves as the communication SINR requirement increases, thereby ensuring that the communication QoS is satisfied. However, the sensing SCNR decreases with higher communication QoS requirements due to the reallocation of resources toward the communication function, which reduces the resources available for sensing. This reduction in SCNR underscores the inherent trade-off between multi-user communication and radar sensing in ISAC systems.

\section{Conclusions}\label{sec:conclusion}
\vspace{0.0 cm}

In this paper, we investigated the potential of leveraging multiple RISs in ISAC systems to realize a novel multi-perspective observation framework. This approach exploits the unique characteristics of target and clutter echo signals across spatial, delay, and Doppler domains. By jointly designing the space-time transmit waveform at the dual-functional BS, the reflection coefficients of active RISs, and the space-time receive filters, the proposed framework aims to maximize the radar output SCNR while ensuring the QoS requirements of communication users. To address the complex and highly non-convex optimization problem, we developed an efficient algorithm that integrates FP MM, and ADMM. Extensive simulation results validate the effectiveness of the proposed multi-perspective observation strategy and highlight the superiority of the joint space-time design over conventional beamforming techniques that focus exclusively on the spatial domain. This work provides a promising avenue for advancing ISAC systems in challenging and clutter-rich environments.



\begin{thebibliography}{99}

\bibitem{LiuFan JSEC 2022} F. Liu, Y. Cui, C. Masouros, J. Xu, T. X. Han, and Y. C. Eldar, ``Integrated sensing and communications: Toward dual-functional wireless networks for 6G and beyond,'' \textit{IEEE J. Sel. Areas Commun.}, vol. 40, no. 6, pp. 1728-1767, Jun. 2022.

\bibitem{Cui Yuanhao Net 2021} Y. Cui, F. Liu, X. Jing, and J. Mu, ``Integrating sensing and communications for ubiquitous IoT: Applications, trends, and challenges,'' \textit{IEEE Netw.}, vol. 35, no. 5, pp. 158-167, Sep. 2021.

\bibitem{Andrew Zhang JSTSP 2021} J. A. Zhang \textit{et al.}, ``An overview of signal processing techniques for joint communication and radar sensing,'' \textit{IEEE J. Sel. Topics Signal Process.}, vol. 15, no. 6, pp. 1295-1315, Nov. 2021.

\bibitem{Liu Fan TCOM 2020} F. Liu, C. Masouros, A. P. Petropulu, H. Griffiths, and L. Hanzo, ``Joint radar and communication design: Applications, state-of-the-art, and the road ahead,'' \textit{IEEE Trans. Commun.}, vol. 68, no. 6, pp. 3834-3862, Jun. 2020.

\bibitem{Liu Fan TSP 2022} F. Liu, Y.-F. Liu, A. Li, C. Masouros and Y. C. Eldar, ``Cram{\'e}r-Rao bound optimization for joint radar-communication Beamforming,'' \textit{IEEE Trans. Signal Process.}, vol. 70, pp. 240-253, 2022.


\bibitem{Liu TSP 2020} X. Liu, T. Huang, N. Shlezinger, Y. Liu, J. Zhou, and Y. C. Eldar, ``Joint transmit beamforming for multiuser MIMO communications and MIMO radar,'' \textit{IEEE Trans. Signal Process.}, vol. 68, pp. 3929-3944, 2020.

\bibitem{Liu Fan TSP 2020} F. Liu, C. Masouros, A. Li, H. Sun, and L. Hanzo, ``MU-MIMO communications with MIMO radar: From co-existence to joint transmission,'' \textit{IEEE Wireless Commun.},vol. 17, no. 4, pp. 2755-2770, Apr. 2018.

\bibitem{Liu Rang JSEC 2022} R. Liu, M. Li, Q. Liu, and A. L. Swindlehurst, ``Joint waveform and filter designs for STAP-SLP-based MIMO-DFRC systems,'' \textit{IEEE J. Sel. Areas Commun.}, vol. 40, no. 6, pp. 1918-1931, Jun. 2022.


\bibitem{Liu Guangyi JSEC 2024} G. Liu, R. Xi, Z. Han, L. Han, and X. Zhang, ``Cooperative sensing for 6G mobile cellular networks: Feasibility, performance, and field trial,'' \textit{IEEE J. Sel. Areas Commun.}, vol. 42, no. 10, pp. 2863-2876, Oct. 2024.


\bibitem{Meng Kaitao MWC 2024} K. Meng, C. Masouros, A. P. Petropulu, and L. Hanzo, ``Cooperative ISAC networks: Opportunities and challenges,'' \textit{IEEE Wireless Commun.}, doi: 10.1109/MWC.008.2400151.


\bibitem{Wei Zhiqing 2024} Z. Wei, H. Liu, H. Li, W. Jiang, Z. Feng, H. Wu, and P. Zhang, ``Integrated sensing and communication enabled cooperative passive sensing using mobile communication system,'' May 2024. [Online]. Available: https://arxiv.org/abs/2405.09179


\bibitem{ElMossallamy TCCN 2020} M. A. ElMossallamy, H. Zhang, L. Song, K. G. Seddik, Z. Han, and G. Y. Li, ``Reconfigurable intelligent surfaces for wireless communications: Principles, challenges, and opportunities,'' \textit{IEEE Trans. Cogn. Commun. Netw.}, vol. 6, no. 3, pp. 990-1002, Sept. 2020.

\bibitem{Basharaty WCOM 2021} S. Basharat, S. A. Hassan, H. Pervaiz, A. Mahmood, Z. Ding, and M. Gidlund, ``Reconfigurable intelligent surfaces: Potentials, applications, and challenges for 6G wireless networks,'' \textit{IEEE Wireless Commun.}, vol. 28, no. 6, pp. 184-191, Dec. 2021.

\bibitem{Renzo JSEC 2020} M. Di Renzo et al., ''Smart radio environments empowered by reconfigurable intelligent surfaces: How it works, state of research, and the road ahead,'' \textit{IEEE J. Sel. Areas Commun.}, vol. 38, no. 11, pp. 2450-2525, Nov. 2020.


\bibitem{Wu Qingqing Cmag 2020} Q. Wu and R. Zhang, ``Towards smart and reconfigurable environment: Intelligent reflecting surface aided wireless network,'' \textit{IEEE Commun.
Mag.}, vol. 58, no. 1, pp. 106-112, Jan. 2020.



\bibitem{Liu Rang WCOM 2023} R. Liu, M. Li, H. Luo, Q. Liu, and A. L. Swindlehurst, ``Integrated sensing and communication with reconfigurable intelligent surfaces: Opportunities, applications, and future directions,'' \textit{IEEE Wireless Commun.}, vol. 30, no. 1, pp. 50-57, Feb. 2023.

\bibitem{Meng Kaitao WCOM 2024} K. Meng, Q. Wu, C. Masouros, W. Chen and D. Li, ``Intelligent surface empowered integrated sensing and communication: From coexistence to reciprocity,'' \textit{IEEE Wireless Commun.}, vol. 31, no. 5, pp. 84-91, Oct. 2024.

\bibitem{Luo Honghao TVT 2023} H. Luo, R. Liu, M. Li, and Q. Liu, ``RIS-aided integrated sensing and communication: Joint beamforming and reflection design,'' \textit{IEEE Trans. Veh. Technol.}, vol. 72, no. 7, pp. 9626-9630, Jul. 2023.

\bibitem{Xu Yongqing TCOM 2024} Y. Xu, Y. Li, J. A. Zhang, M. D. Renzo, and T. Q. S. Quek, ``Joint beamforming for RIS-assisted integrated sensing and communication systems,'' \textit{IEEE Trans. Commun.}, vol. 72, no. 4, pp. 2232-2246, Apr. 2024.

\bibitem{Liu Rang JSTSP 2022} R. Liu, M. Li, Y. Liu, Q. Wu, and Q. Liu, ``Joint transmit waveform and passive beamforming design for RIS-aided DFRC systems,'' \textit{IEEE J. Sel. Topics Signal Process. (JSTSP)}, vol. 16, no. 5, pp. 995-1010, Aug. 2022.

\bibitem{Yuan Fang TCOM 2024} Y. Fang, S. Zhang, X. Li, X. Yu, J. Xu and S. Cui, ``Multi-IRS-enabled integrated sensing and communications,'' \textit{IEEE Trans. Commun.}, vol. 72, no. 9, pp. 5853-5867, Sept. 2024.

\bibitem{Wei Tong ICOM 2023} T. Wei, L. Wu, K. V. Mishra, and M. R. B. Shankar, ``Multi-IRS-aided Doppler-tolerant wideband DFRC system,'' \textit{IEEE Trans. Commun.}, vol. 71, no. 11, pp. 6561-6577, Nov. 2023.


\bibitem{Zhang Zijian TCOM 2023} Z. Zhang \textit{et al.}, ``Active RIS vs. passive RIS: Which will prevail in 6G?'' \textit{IEEE Trans. Commun.}, vol. 71, no. 3, pp. 1707-1725, Mar. 2023.

\bibitem{Long Ruizhe TWC 2021} R. Long, Y.-C. Liang, Y. Pei, and E. G. Larsson, ``Active reconfigurable intelligent surface-aided wireless communications,'' \textit{IEEE Trans. Wireless Commun.}, vol. 20, no. 8, pp. 4962-4975, Aug. 2021.


\bibitem{Zhu Qi TVT 2023} Q. Zhu, M. Li, R. Liu, and Q. Liu, ``Joint transceiver beamforming and reflecting design for active RIS-aided ISAC systems,'' \textit{IEEE Trans. Veh. Technol.}, vol. 72, no. 7, pp. 9636-9640, July 2023.

\bibitem{Yu Zhiyuan TCOM 2024} Z. Yu \textit{et al.}, ``Active RIS-aided ISAC systems: Beamforming design and performance analysis,'' \textit{IEEE Trans. Commun.}, vol. 72, no. 3, pp. 1578-1595, Mar. 2024.

\bibitem{Zhu Qi TWC 2024} Q. Zhu, M. Li, R. Liu, and Q. Liu, ``Cram{\'e}r-Rao bound optimization for active RIS-empowered ISAC systems,'' \textit{IEEE Trans. Wireless Commun.}, vol. 23, no. 9, pp. 11723-11736, Sept. 2024.

\bibitem{Zhang Yang TVT 2024} Y. Zhang \textit{et al.}, ``Secure wireless communication in active RIS-assisted DFRC systems,'' \textit{IEEE Trans. Veh. Technol.}, doi: 10.1109/TVT.2024.3438151.


\bibitem{Liu Mengyu TCCN 2024} M. Liu \textit{et al.}, ``Joint beamforming design for double active RIS-assisted radar-communication coexistence systems,'' \textit{IEEE Trans. Cogn. Commun. Networking}, vol. 10, no. 5, pp. 1704-1717, Oct. 2024.


\bibitem{Wang Fangzhou SPL 2022} F. Wang, H. Li and J. Fang, ``Joint active and passive beamforming for IRS-assisted radar,'' \textit{IEEE Signal Process. Lett.}, vol. 29, pp. 349-353, 2022.



\bibitem{Zuo Lei TAES 2024} L. Zuo, T. Yang, X. Lu, M. Zhang, and Z. Lan, ``Joint design of transmit waveforms and receive filters for MIMO-STAP radar with two sequential constraints,'' \textit{IEEE Trans. Aerosp. Electron. Syst.}, vol. 60, no. 4, pp. 5035-5048, Aug. 2024.


\bibitem{Tang Bo TSP 2024} B. Tang and J. Tang, ``Joint design of transmit waveforms and receive filters for MIMO radar space-time adaptive processing,'' \textit{IEEE Trans. Signal Process.}, vol. 64, no. 18, pp. 4707-4722, Sept. 2016.



\bibitem{Xie Zhuang TSP 2024} Z. Xie, L. Wu, J. Zhu, M. Lops, X. Huang and M. R. B. Shankar, ``RIS-aided radar for target detection: Clutter region analysis and joint active-passive design,'' \textit{IEEE Trans. Signal Process.}, vol. 72, pp. 1706-1723, 2024.

\bibitem{MA ICST 2018} M. Alodeh, \textit{et al.}, ``Symbol-level and multicast precoding for multiuser multiantenna downlink: A state-of-art, classification, and challenges,'' \textit{IEEE Commun. Surveys  Tut.}, vol. 20, no. 3, pp. 1733-1757, May 2018.


\bibitem{Li ICST 2020} A. Li, \textit{et. al.}, ``A tutorial on interference exploitation via symbol-level precoding: Overview, state-of-the-art and future directions,'' \textit{IEEE Commun. Surveys Tut.}, vol. 22, no. 2, pp. 796-839, 2nd quarter 2020.


\bibitem{Liu TWC 2021} R. Liu, M. Li, Q. Liu, and A. L. Swindlehurst, ``Joint symbol-level precoding and reflecting designs for IRS-enhanced MU-MISO systems,'' \textit{IEEE Trans. Wireless Commun.}, vol. 20, no. 2, pp. 798-811, Feb. 2021.


\bibitem{Horn 1990} R. A. Horn and C. R. Johnson, \textit{Matrix Analysis}. Cambridge, U.K.: Cambridge Univ. Press, 1990.


\end{thebibliography}
\end{document}